\documentclass[twocolumn,floatfix,superscriptaddress,prd]{revtex4}
\usepackage{subfigure,dcolumn,jabbrv}
\usepackage{graphicx}% Include figure files
\usepackage{dcolumn}% Align table columns on decimal point
\usepackage{bm}% bold math
\usepackage{CJK}
\usepackage{url}
\usepackage{color}
\usepackage{booktabs}
\usepackage{natbib}

\usepackage{amsmath,bm}
\usepackage[utf8]{inputenc}
\usepackage[british,UKenglish,USenglish,american]{babel}
\usepackage{bbm}
\usepackage{amssymb}
\usepackage{amsmath}
\usepackage{hyperref}
\usepackage{cleveref}
\usepackage{verbatim}

\usepackage{graphicx}

\usepackage[normalem]{ulem}
\usepackage{appendix}
\usepackage{color}

\usepackage{pdfpages}
\usepackage{chngpage}
\newcommand{\dd}{\mathrm{d}}

\usepackage{pgf}

\usepackage{array}

\begin{document}

%\title{Bayesian Analysis of the Nuclear Matter EOS with Crossover Phase Transition}
%\title{Bayesian Inference of the Nuclear Matter Equation of State with a Smooth Crossover Between Hadronic and Quark Phases from Neutron Star Observations}
\title{Bayesian Constraints on the Neutron Star Equation of State with a Smooth Hadron–Quark Crossover}

\author{Xavier Grundler\footnote{xgrundler@leomail.tamuc.edu}}
\affiliation{Department of Physics and Astronomy, East Texas A$\&$M University, Commerce, TX 75429, USA}
\author{Bao-An Li\footnote{Corresponding Author: Bao-An.Li@etamu.edu}}
\affiliation{Department of Physics and Astronomy, East Texas A$\&$M University, Commerce, TX 75429, USA}
\date{\today}

\begin{abstract}
We perform a Bayesian inference of the dense-matter equation of state (EOS) within a unified framework that incorporates hadronic matter, quark matter, and a smooth hadron–quark crossover. The EOS is constrained using physical consistency conditions, gravitational-wave data from GW170817, NICER mass–radius measurements, and hypothetical future high-precision radius observations. In contrast to most previous studies that assume a sharp first-order phase transition or fix part of the EOS, we simultaneously infer hadronic, quark, and crossover parameters within a single statistical framework.
We find that current observations strongly constrain the density dependence of the nuclear symmetry energy, particularly its slope and curvature, while the highest-density hadronic parameters and quark-matter properties remain only weakly constrained. The posterior distributions favor a crossover centered at an energy density $\varepsilon \sim (4$–$6)\varepsilon_0$ with a width $\Gamma \sim (0.5$–$1.0)\varepsilon_0$. A pronounced peak in the speed of sound emerges naturally near the crossover region, typically around $4\varepsilon_0$, and often coincides with the central densities of $\sim 2\,M_\odot$ neutron stars. We further show that the trace anomaly exhibits a remarkably universal behavior across the accepted EOS ensemble and remains largely insensitive to current observational constraints. This indicates that present data primarily probe the low-to-intermediate density EOS, while robust inference of quark matter and genuinely high-density physics will require next-generation precision radius measurements or complementary observables.
\end{abstract}

\maketitle

\section{Introduction}

Exploring the quantum chromodynamic (QCD) phase diagram is a central problem in both astrophysics and nuclear physics; see, e.g., Refs.~\cite{Bogdanov:2022faf, Sorensen:2023zkk} for recent reviews. At finite temperature and vanishing baryon density, hadronic matter (HM) is known to undergo a smooth crossover transition to deconfined quark matter (QM) \cite{Fukushima:2025ujk}. However, the existence and location of one or more critical endpoints---separating regions of first-order phase transitions from crossover behavior in which the two phases become indistinguishable---remain unknown; see, e.g., Refs.~\cite{Ecker:2025vnb, Ferreira:2018sun, Stephanov:2024xkn, Ferroni:2010ct, Blaschke:2013ana}. While terrestrial experiments at heavy-ion facilities probe the QCD phase diagram at high temperatures and finite baryon densities up to a few times nuclear saturation density \cite{Du:2024wjm, Lattimer:2023rpe}, it is not possible to create large volumes of stable, cold, dense nuclear matter in the laboratory. Neutron stars (NS), by contrast, are effectively at zero temperature and can reach central densities of up to six to ten times saturation density, making them unique astrophysical laboratories for exploring the QCD phase diagram in the high-density, low-temperature regime.

In principle, the properties of dense nuclear matter could be derived directly from QCD. In practice, however, lattice QCD is limited to vanishing baryon density, while perturbative QCD is reliable only at asymptotically high densities, far beyond those realized in neutron stars. As a result, a wide range of theoretical approaches have been developed to model the nuclear matter equation of state (EOS), including mean-field theories, effective field theories, Gaussian processes, and phenomenological parameterizations. The EOS, defined as the pressure as a function of energy density, $P(\varepsilon)$, uniquely determines the neutron star mass--radius (MR) relation through the Tolman--Oppenheimer--Volkoff (TOV) equations \cite{tolman1939, oppenheimer1939}. Consequently, precise measurements of neutron star masses and radii provide direct constraints on the EOS. In this work, we exploit this connection through a Bayesian analysis informed by modern neutron star observations, including the results with the smallest reported errors from the Neutron Star Interior Composition Explorer (NICER).

To construct the EOS, we adopt a phenomenological framework that couples a six-parameter HM meta-model to a two-parameter QM model via a two-parameter smooth crossover function. In many previous studies, including several of our own, the transition between HM and QM has been modeled as a first-order phase transition using either Maxwell or Gibbs constructions; see, e.g., Refs.~\cite{Brandes:2023bob, Alarcon:2025qmz} and references therein. However, lattice QCD results at vanishing baryon density \cite{Fukushima:2025ujk}, together with implications from recent astrophysical measurements \cite{Brandes:2023hma}, motivate exploring the impact of replacing a sharp first-order transition with a smooth crossover. A recent Bayesian analysis using NICER data found slight evidence for a crossover, based on comparisons of the mass, radius, and fundamental mode between purely nucleonic neutron stars and those with a crossover \cite{Roy:2024sjx}. The role of smooth crossovers in neutron stars has also been investigated in a variety of contexts, including quasinormal modes \cite{Constantinou:2021hba, Pradhan:2023zmg, Sotani:2023zkk}, observational signatures \cite{Fujimoto:2024ymt, Dexheimer:2014pea}, peaks in the speed of sound \cite{Tajima:2024qzj, Iida:2022hyy}, and the emergence of twin-star solutions \cite{Ayriyan:2017nby, Abgaryan:2018gqp, Alvarez-Castillo:2014dva, Blaschke:2020qqj}. From a physical perspective, a crossover scenario is supported by the possible percolation of quarks among hadrons at high densities \cite{Masuda:2012ed, Kojo:2014rca, Takatsy:2023xzf} or by the existence of quarkyonic matter \cite{McLerran:2018hbz, McLerran:2020rnw}.

In this study, we employ a broad ensemble of EOSs together with a new trace-anomaly-based parameterization for quark matter. Within a Bayesian framework, we infer the most probable values of the EOS parameters and quantify their uncertainties using current neutron star observational data. We place particular emphasis on the squared speed of sound, $c_s^2(\varepsilon)$, and the trace anomaly, defined as $\Delta = 1/3 - P/\varepsilon$, which characterize the stiffness of dense matter and the degree of conformal symmetry breaking, respectively. The energy-density dependence of these quantities plays a crucial role in shaping the EOS and has a direct impact on neutron star global properties.

In contrast to previous studies, the present work introduces several key advances. 
First, we perform a fully unified Bayesian inference of the dense-matter EOS in which 
hadronic, quark, and crossover parameters are simultaneously constrained, rather than 
fixing one sector. Second, we employ a 
trace-anomaly-based parameterization of quark matter, enabling a direct and largely 
model-independent assessment of deviations from conformal behavior. Third, we show 
that a smooth hadron–quark crossover generically induces a peak in the speed of sound, 
whose location correlates strongly with the crossover density, thereby providing a 
physical link between microscopic EOS structure and neutron star observables. Finally, 
we systematically examine the impact of data selection and demonstrate that current 
observations robustly constrain the symmetry energy at low-to-intermediate densities, 
while leaving the quark-matter sector largely unconstrained. While previous studies have explored crossover constructions or Bayesian inference separately, the present work combines both within a single, statistically consistent framework that treats all EOS sectors on equal footing.

The remainder of this paper is organized as follows. In Sec.~\ref{met}, we describe our EOS model and outline the Bayesian methodology. Our main results are presented and discussed in Sec.~\ref{results}. We conclude with a summary and outlook in Sec.~\ref{summary}.

\section{Methods}\label{met}
Below, we describe both the EOS meta-model and the Bayesian framework adopted in this work. 
The HM EOS employed here is the same as that used in our previous studies 
\cite{zhang2018combined, Zhang:2019fog, Zhang:2021xdt, xie2019bayesian, xie2020bayesian, xie2021bayesian, Zhang:2023wqj, Zhang:2024npg, Xie:2024mxu, Li:2024imk, LiEPJA, universeReview, hybridPrecision, Grundler:2025mcz}. 
Our Bayesian scheme is also closely aligned with these earlier analyses.

\subsection{NS EOS Meta-Model}
Our NS EOS consists of four components, three of which are described by meta-models. 
For the low-density crust, we adopt the Negele--Vautherin (NV) EOS \cite{negele1973neutron} and the Baym--Pethick--Sutherland (BPS) EOS \cite{baym1971ground} for the inner and outer crusts, respectively. 
These are connected to the HM core meta-model at the density where the uniform HM EOS becomes thermodynamically unstable, following Refs.~\cite{lattimer:2006xb, kubis2007nuclear, Xu:2009vi}. 
Finally, we parameterize a smooth crossover region that interpolates between the HM and QM EOSs. 
The detailed parameterizations are described in the following subsections. 
Unless otherwise specified, we use natural units with $c=1$.

\subsubsection{HM EOS}
The HM meta-model is based on a parameterization of the binding energy per nucleon of 
$\beta$-equilibrated $npe\mu$ matter,
\begin{equation}\label{bindingenergy}
E(\rho,\delta) = E_0(\rho) + E_{\rm sym}(\rho)\,\delta^2 + \mathcal{O}(\delta^4),
\end{equation}
where we adopt the empirical isospin-parabolic approximation for neutron-rich matter 
\cite{bombaci1991asymmetric}. 
Here, $\rho$ denotes the baryon number density and 
$\delta = (\rho_n - \rho_p)/\rho$ is the isospin asymmetry, with $\rho_n$ and $\rho_p$ being the neutron and proton densities, respectively. 
The first term, $E_0(\rho)$, is the EOS of symmetric nuclear matter (SNM), while 
$E_{\rm sym}(\rho)$ is the nuclear symmetry energy, which quantifies the energy cost of converting protons into neutrons \cite{drLiSymEnerTalk}. 

Both $E_0(\rho)$ and $E_{\rm sym}(\rho)$ are expanded around the saturation density $\rho_0$ as
\begin{eqnarray}\label{E0para}
E_{0}(\rho) &=& E_0(\rho_0)
+ \frac{K_0}{2}\left(\frac{\rho-\rho_0}{3\rho_0}\right)^2
+ \frac{J_0}{6}\left(\frac{\rho-\rho_0}{3\rho_0}\right)^3,\\
E_{\rm sym}(\rho) &=& E_{\rm sym}(\rho_0)
+ L\left(\frac{\rho-\rho_0}{3\rho_0}\right)
+ \frac{K_{\rm sym}}{2}\left(\frac{\rho-\rho_0}{3\rho_0}\right)^2 \nonumber\\
&+& \frac{J_{\rm sym}}{6}\left(\frac{\rho-\rho_0}{3\rho_0}\right)^3,
\label{Esympara}
\end{eqnarray}
where the expansion coefficients are treated as free parameters in the Bayesian analysis. 
Thus, the series should be viewed as a flexible parameterization rather than a strict Taylor expansion, and convergence at suprasaturation densities is not required. 
In practice, the highest-order coefficients $J_0$ and $J_{\rm sym}$ encode the effective contributions from all higher-order terms.

The coefficients are defined by
\begin{eqnarray}
K_0 &=& 9\rho_0^2\left.\frac{\partial^2 E_0}{\partial\rho^2}\right|_{\rho_0},\\
J_0 &=& 27\rho_0^3\left.\frac{\partial^3 E_0}{\partial\rho^3}\right|_{\rho_0},\\
L &=& 3\rho_0\left.\frac{\partial E_{\rm sym}}{\partial\rho}\right|_{\rho_0},\\
K_{\rm sym} &=& 9\rho_0^2\left.\frac{\partial^2 E_{\rm sym}}{\partial\rho^2}\right|_{\rho_0},\\
J_{\rm sym} &=& 27\rho_0^3\left.\frac{\partial^3 E_{\rm sym}}{\partial\rho^3}\right|_{\rho_0},
\end{eqnarray}
which correspond to the curvature and skewness of SNM, and the slope, curvature, and skewness of the symmetry energy, respectively, all evaluated at $\rho_0$. 
We fix $\rho_0 = 0.16~\mathrm{fm}^{-3}$ (the corresponding energy density of SNM is $\varepsilon_0 \simeq 150~\mathrm{MeV/fm}^3$) and 
$E_0(\rho_0) = -16$ MeV, which are well constrained by terrestrial nuclear experiments.

The pressure is obtained from the thermodynamic relation
\begin{equation}
P_{\rm HM}(\rho,\delta) 
= \rho^2 \frac{d}{d\rho}\left(\frac{\varepsilon_{\rm HM}(\rho,\delta)}{\rho}\right),
\end{equation}
where the HM energy density is
$\varepsilon_{\rm HM}(\rho,\delta) = \rho [E(\rho,\delta) + M_N] + \varepsilon_l(\rho,\delta)$. 
Here, $M_N$ is the average nucleon mass, and $\varepsilon_l$ denotes the lepton energy density, which is calculated using a non-interacting relativistic Fermi gas \cite{oppenheimer1939}.

\subsubsection{QM EOS}
For the QM meta-model, we adopt the \texttt{p1} parameterization introduced in Ref.~\cite{caili_newpara}, which does not assume a specific microscopic composition of quark matter, such as the presence or absence of strange quarks. 
This parameterization is formulated in terms of the trace anomaly \cite{Fujimoto:2024ymt},
\begin{equation}\label{traceanomaly}
\Delta \equiv \frac{1}{3} - \frac{P}{\varepsilon}.
\end{equation}
Following Ref.~\cite{caili_newpara}, we parameterize $\Delta(\varepsilon)$ as
\begin{equation}\label{p1}
\Delta = \frac{1}{3}\left(1 - f\, t\, \varepsilon_*^{a}\right)
e^{-t\varepsilon_*^{a}},
\end{equation}
where $t$ and $a$ are model parameters, 
and $\varepsilon_* \equiv \varepsilon/\varepsilon_0$. 
The constant $f$ is fixed to $f = 1.0318$ such that the minimum value of the trace anomaly is 
$\Delta_{\rm min} \simeq -0.048$, which represents a lower bound motivated by general relativistic considerations \cite{caili_newpara}. 
By construction, this parameterization approaches the conformal limit $c_s^2 \to 1/3$ as $\varepsilon \to \infty$, consistent with perturbative QCD expectations 
\cite{Somasundaram:2022ztm, Zhou:2023zrm}. 
The QM EOS $P_{\rm QM}(\varepsilon)$ follows directly from the definition of $\Delta$.

In terms of the trace anomaly, the squared speed of sound is given by \cite{Fujimoto:2024ymt}
\begin{equation}
c_s^2(\varepsilon)
\equiv \frac{dP}{d\varepsilon}
= -\varepsilon_* \frac{d\Delta}{d\varepsilon_*}
+ \frac{1}{3} - \Delta,
\label{ss_decom}
\end{equation}
which shows explicitly that, depending on the values of $a$ and $t$, the parameterization in Eq.~(\ref{p1}) can generate a pronounced peak in the sound-speed profile. 
Such a peak is a common feature of many theoretical models and is often required to sufficiently stiffen the EOS in order to support massive neutron stars consistent with current observations 
\cite{Bedaque:2014sqa, Marczenko:2024uit, Cai:2023pkt, Cai:2023ajw}.

\subsubsection{Crossover Region}
In contrast to much of our previous work, we employ a smooth crossover, rather than a sharp first-order phase transition, to connect the HM and QM EOSs. 
A variety of interpolation schemes have been proposed in the literature. 
Common approaches include polynomial interpolations in chemical potential $\mu$ 
\cite{Baym:2017whm, Baym:2019iky, Kojo:2014rca, Kojo:2021wax, Minamikawa:2020jfj, Takatsy:2023xzf, Blaschke:2020qqj}, 
piecewise-polynomial methods \cite{Ayriyan:2017nby, Ayriyan:2021prr}, 
and switching functions based on exponential forms 
\cite{Constantinou:2021hba, Lavagno:2025xwl, Kapusta:2021ney, Blaschke:2021poc} 
or hyperbolic tangent functions 
\cite{Hell:2014xva, Yang:2023duo, Qin:2023zrf, Kovacs:2021ger}. 
The EOS introduced in Ref.~\cite{Masuda:2012ed}, for example, employed a hyperbolic tangent function motivated by quark percolation in hadronic matter.

The term ``smooth crossover'' is used somewhat loosely in the literature. 
As noted in Ref.~\cite{Abgaryan:2018gqp}, it is often applied to any transition other than a Maxwell construction, which features a discontinuity in energy density at fixed pressure. 
Here, we classify the order of the phase transition by the continuity properties of the squared speed of sound and its derivatives. 
Specifically, a discontinuity in $c_s^2(\varepsilon)$ corresponds to a first-order transition, while a discontinuity in its derivative corresponds to a second-order transition. 
Polynomial interpolations can only enforce smoothness up to a finite order, due to the limited number of adjustable coefficients. 
To achieve an infinitely differentiable crossover, we therefore adopt a switching function.

Following Ref.~\cite{Hell:2014xva}, the EOS with a crossover is defined as
\begin{eqnarray}
P(\varepsilon) &=& P_{\rm HM}(\varepsilon) f_-(\varepsilon) 
+ P_{\rm QM}(\varepsilon) f_+(\varepsilon),\label{pcross}\\
f_\pm(\varepsilon) &=& \frac{1}{2}\left[1 \pm 
\tanh\!\left(\frac{\varepsilon - \overline{\varepsilon}}{\Gamma}\right)\right],
\label{switchfunc}
\end{eqnarray}
where $\overline{\varepsilon}$ denotes the center of the crossover region and $\Gamma$ controls its width. 
This form differs from Ref.~\cite{Masuda:2012ed} in that energy density, rather than baryon number density, is used as the interpolation variable. 
By construction, the HM EOS dominates for 
$\varepsilon \lesssim \overline{\varepsilon} - 2\Gamma$, 
the QM EOS dominates for 
$\varepsilon \gtrsim \overline{\varepsilon} + 2\Gamma$, 
and the two are smoothly mixed in between.

The squared speed of sound for the crossover EOS is then given by
\begin{eqnarray}
c_s^2(\varepsilon) &=& c_{\rm HM}^2(\varepsilon) f_-(\varepsilon)
+ P_{\rm HM}(\varepsilon)\frac{df_-}{d\varepsilon} \nonumber\\
&+& c_{\rm QM}^2(\varepsilon) f_+(\varepsilon)
+ P_{\rm QM}(\varepsilon)\frac{df_+}{d\varepsilon}.
\label{sss}
\end{eqnarray}

\subsection{Bayesian Analysis}
Bayes' theorem states that
\begin{equation}
    P(\mathcal{M}|D) 
    = \frac{P(D|\mathcal{M})\,P(\mathcal{M})}
    {\int P(D|\mathcal{M})\,P(\mathcal{M})\,d\mathcal{M}},
\end{equation}
which allows us to compute the posterior probability $P(\mathcal{M}|D)$ of a model $\mathcal{M}$ given data $D$, from the likelihood $P(D|\mathcal{M})$ and the prior $P(\mathcal{M})$. The denominator is a normalization constant (the Bayesian evidence).

\begin{table}[htbp]
    \caption{Prior ranges for the EOS parameters.}
    \centering
    \begin{tabular}{ccc}
        \hline\hline
        Parameters &~~~~~Lower limit &~~~~~Upper limit \\
        \hline
        $K_0$ (MeV) & 220 & 260\\
        $J_0$ (MeV) & -400 & 400\\
        $E_{\rm sym}(\rho_0)$ (MeV) & 28.5 & 34.9\\
        $L$ (MeV) & 30 & 90\\
        $K_{\rm sym}$ (MeV) & -400 & 100\\
        $J_{\rm sym}$ (MeV) & -200 & 800\\
        $a$ & 0.6 & 2.42\\
        $t$ & 0.025 & 0.115\\
        $\overline{\varepsilon}$ (MeV) & 300 & 900\\
        $\Gamma$ (MeV) & 5 & 400\\
        \hline\hline
    \end{tabular}
    \label{tab:priors}
\end{table}

\begin{table*}[htbp]
    \caption{NICER data with symmetrized uncertainties used to construct bivariate normal likelihoods, showing $\pm1\sigma$. The value $\rho_{MR}$ is the correlation coefficient.}
    \centering
    \setlength{\tabcolsep}{1em}
    {\renewcommand{\arraystretch}{1.5}
    \begin{tabular}{lccccc}
        \hline\hline
        Name & Mass ($M_\odot$) & Radius (km) & $\rho_{MR}$ & Model & Analysis \\
        \hline
        PSR J0740+6620 & $2.073 \pm 0.069$ & $12.49 \pm 1.08$ & 0.272 & ST-U & Salmi et al. (2024) \cite{Salmi:2024aum, salmi_2024_10519473}\\
        PSR J0030+0451 & $1.40 \pm 0.125$ & $11.71 \pm 0.855$ & 0.878 & ST + PDT & Vinciguerra et al. (2024) \cite{Vinciguerra:2023qxq, vinciguerra_2023_8239000}\\
        PSR J0437+4715 & $1.418 \pm 0.037$ & $11.36 \pm 0.79$ & 0.261 & CST + PDT & Choudhury et al. (2024) \cite{Choudhury:2024xbk, choudhury_2024_13766753}\\
        PSR J0614+3329 & $1.44 \pm 0.065$ & $10.29 \pm 0.935$ & 0.376 & ST + PDT & Mauviard et al. (2025) \cite{Mauviard:2025dmd, mauviard_2025_17380576}\\
        PSR J1231+1411 & $1.04 \pm 0.04$ & $12.6 \pm 0.32$ & 0.0949 & PDT-U & Salmi et al. (2024) \cite{Salmi:2024bss, salmi_2024_13358349}\\
        \hline\hline
    \end{tabular}}
    \label{tab:nicerdata}
\end{table*}

The prior bounds on our model parameters are listed in Table~\ref{tab:priors}. 
For the HM parameters, we sample uniformly and adopt flat priors, treating all values within the specified ranges as equally probable. 
In previous work, we examined the impact of using Gaussian priors for these parameters and found that the resulting posteriors changed only marginally \cite{Xie:2020kta}, albeit in a slightly different context. 
We therefore retain uniform priors here, as they represent the least informative choice.

We adopt the same uniform-prior strategy for the crossover parameters $\overline{\varepsilon}$ and $\Gamma$. 
To our knowledge, no previous studies have systematically explored or constrained prior ranges for these parameters.
For comparison, Ref.~\cite{Hell:2014xva} adopted $\overline{\varepsilon} = 800$~MeV and $\Gamma = 300$~MeV, corresponding to a transition region at several times the saturation density.

For the QM parameters, we employ Gaussian sampling with Gaussian priors centered at 
$\mu_a = 1.51$ with $\sigma_a = 0.302$, and 
$\mu_t = 0.07$ with $\sigma_t = 0.014$. 
These priors are motivated by the analysis of central trace anomalies for 17 NS instances (some of them are for the same NSs but with radii from different measurements or analyses), inferred from their compactness constraints in Ref.~\cite{caili_newpara}.

Our likelihood function is defined as
\begin{equation}
    P(D|\mathcal{M}) 
    = P_{\rm filter} \times P_{\rm mass,max} \times P_R.
\end{equation}
The first term, $P_{\rm filter}$, enforces basic physical consistency. 
EOSs that violate causality are rejected; EOSs that become mechanically unstable are truncated at the point where $dP/d\varepsilon < 0$; and EOSs must yield a positive crust--core transition pressure. 
In addition to these standard filters used in our previous studies, we impose two further conditions on the crossover region. 
First, following Ref.~\cite{Masuda:2012ed}, we require the crossover region, defined by 
$\overline{\varepsilon} - 2\Gamma < \varepsilon < \overline{\varepsilon} + 2\Gamma$, 
to begin above the saturation density, since nuclear matter at saturation is known to be purely hadronic. 
Second, we require the HM pressure to remain positive throughout the crossover region; although the HM contribution may decrease as the QM component grows, it should not become negative while still contributing to the total EOS.

The factor $P_{\rm mass,max}$ is implemented as a step function requiring that the EOS support a neutron star with a maximum mass of at least $1.97\,M_\odot$ \cite{Antoniadis:2013pzd}. This choice provides a conservative lower bound on the maximum nonrotating mass $M_{\rm TOV}$. Although higher-mass measurements exist—for example, PSR J0952+0607 with $M = 2.35 \pm 0.17\,M_\odot$ \cite{Romani:2022jhd}—we adopt this conservative threshold with a sharp cutoff to avoid additional uncertainties associated with observational modeling and rapid rotation. For a detailed discussion of the impact of different choices of $P_{\rm mass,max}$ (including alternative cutoffs or distributions) in Bayesian analyses of neutron star data, see Sec.~4.4 of Ref.~\cite{xie2019bayesian}.

The final term, $P_R$, incorporates the observational mass--radius data. 
When the mass is assumed to be known exactly, we use a univariate Gaussian likelihood,
\begin{equation}
    P_R = \prod_{j=1}^{N}
    \frac{1}{\sqrt{2\pi}\sigma_{{\rm obs},j}}
    \exp\!\left[
    -\frac{(R_{\rm th}-R_{{\rm obs},j})^2}
    {2\sigma_{{\rm obs},j}^2}
    \right],
\end{equation}
where $R_{\rm th}$ is the theoretical radius obtained from solving the TOV equations, and $R_{{\rm obs},j}$ and $\sigma_{{\rm obs},j}$ are the observed radius and uncertainty for the $j$-th source.

When both mass and radius have observational uncertainties, we adopt a bivariate normal likelihood following Ref.~\cite{Blaschke:2020qqj} as an approximation of the true, asymmetric distribution of the data,
\begin{equation}\label{bivariate}
    \frac{1}{2\pi \sigma_M\sigma_R \sqrt{1 - \rho_{MR}^2}}
    \exp\!\left[-\frac{x}{2(1 - \rho_{MR}^2)}\right],
\end{equation}
with
\begin{equation}\label{xdef}
    x = \frac{(M - \mu_M)^2}{\sigma_M^2}
    - 2\rho_{MR}\frac{(M - \mu_M)(R - \mu_R)}{\sigma_M\sigma_R}
    + \frac{(R - \mu_R)^2}{\sigma_R^2}.
\end{equation}
Here, $(M,R)$ are the theoretical mass--radius pairs from the TOV equations, 
$(\mu_M,\sigma_M,\mu_R,\sigma_R)$ are the observational values, and 
$\rho_{MR}$ is the correlation coefficient. 
For each observation, we marginalize the likelihood as done in Ref. \cite{Blaschke:2020qqj}, and take the product over all sources.

\begin{figure}
    \centering
    \includegraphics[width=\linewidth]{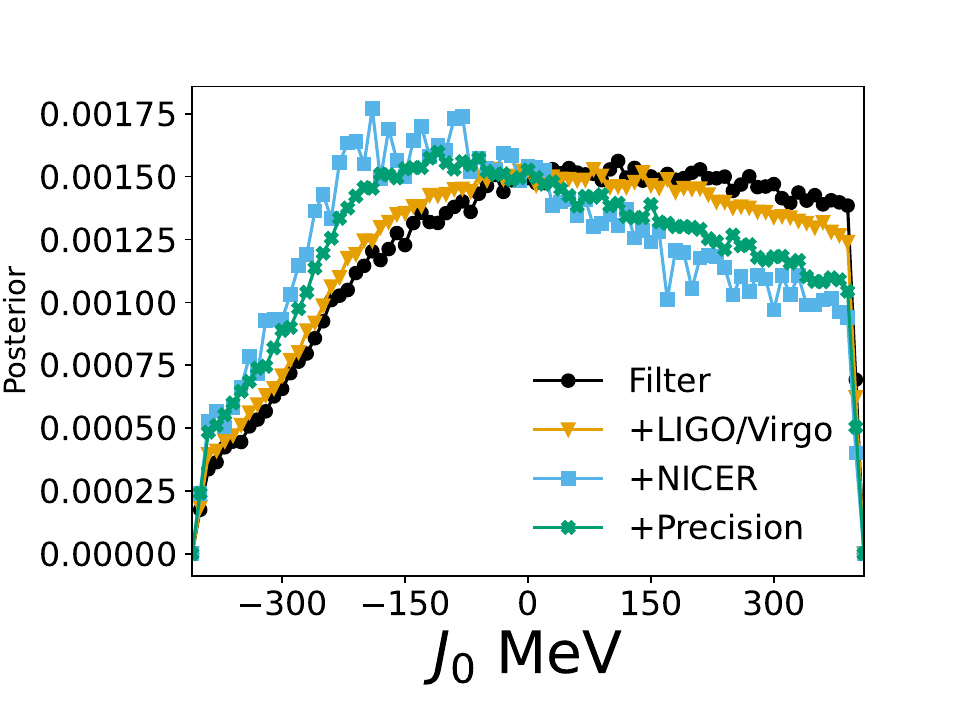}

    \vspace{-1cm}
    
    \includegraphics[width=\linewidth]{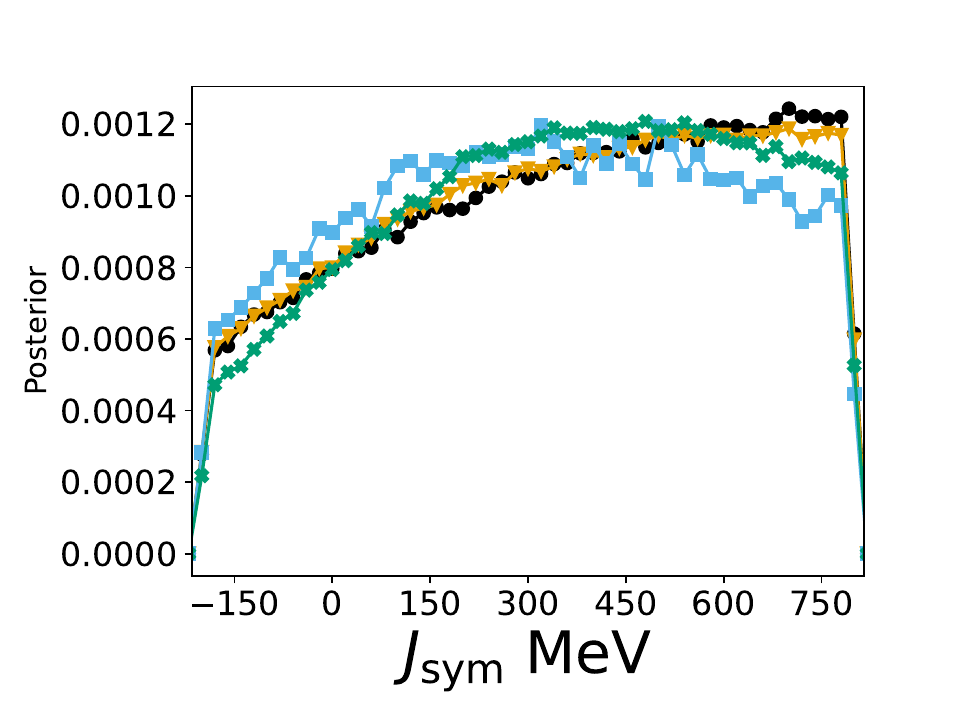}

    \vspace{-1cm}
    
    \includegraphics[width=\linewidth]{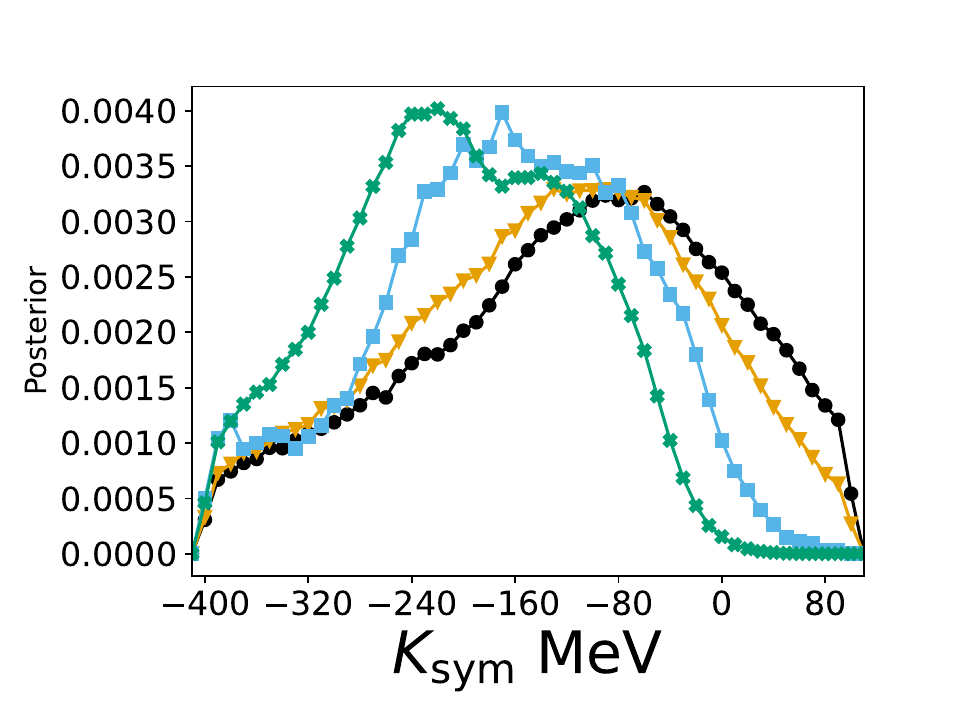}

    \vspace{-1cm}
    
    \includegraphics[width=\linewidth]{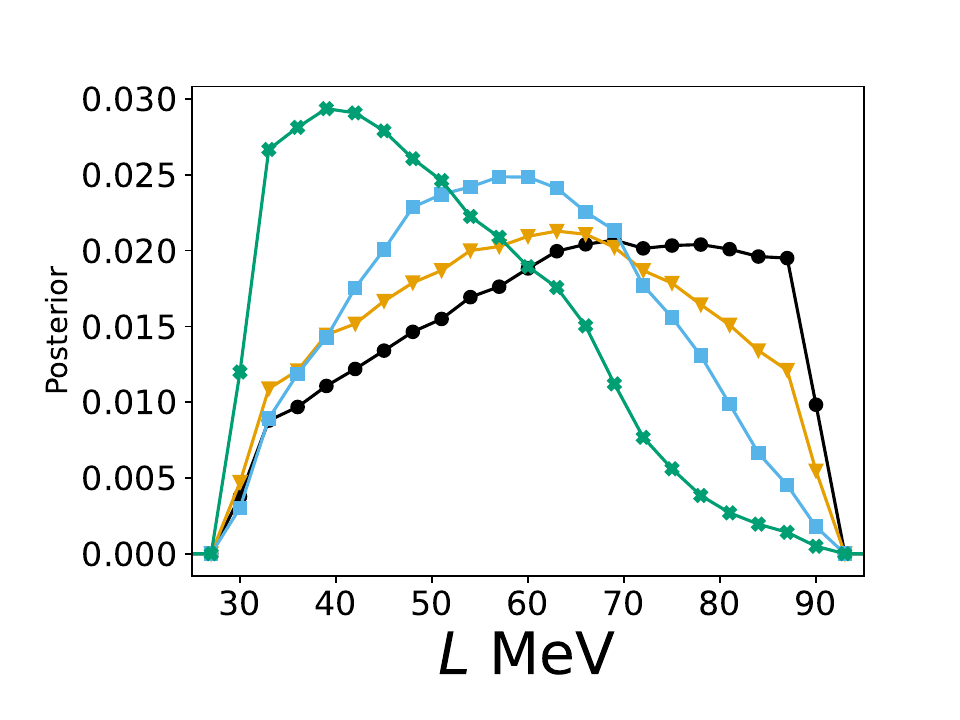}

    \vspace{-0.5cm}
    \caption{PDFs of the HM parameters.}
    \label{fig:hmpara}
\end{figure}

We consider four observational scenarios to quantify the impact of different observing methods:
\begin{enumerate}
    \item $P_R = 1$: no observational data, which quantifies the constraining power of fundamental physical requirements alone.
    
    \item $R_{1.4} = 11.9 \pm 1.4$~km at 90\% confidence level (corresponding to $\sigma = 0.875$ km) from the LIGO/Virgo analysis of GW170817 \cite{abbott2018gw170817}, representative of current typical constraints.
    
    \item The NICER measurements with the smallest reported errors summarized in Table~\ref{tab:nicerdata}.
    
    \item $R_{1.4} = 11.9 \pm 0.2$~km, corresponding to a hypothetical future measurement with significantly improved precision using third-generation gravitational wave detectors \cite{Chatziioannou:2021tdi, Pacilio:2021jmq, Bandopadhyay:2024zrr, Finstad:2022oni, Walker:2024loo}.
\end{enumerate}

For PSR J0740+6620, PSR J0030+0451, and PSR J0437+4715, multiple NICER analyses exist based on different hotspot models and data selections 
\cite{Riley:2021pdl, Miller:2021qha, Dittmann:2024mbo, Riley:2019yda, Miller:2019cac, Miller:2025qfq}. 
The choices in Table~\ref{tab:nicerdata} are not intended to identify a preferred analysis, but rather to use the results with the smallest reported errors to illustrate the impact of incorporating multiple observational constraints. 
Similarly, the results for PSR J1231+1411 are considered less robust due to model convergence issues \cite{Salmi:2024bss}, but are included to provide constraints at low mass and to ensure broad coverage of the mass--radius relation. We use the analyses with the smallest report errors, even if they are considered less realistic, to determine the effect of many high-precision x-ray measurements. This provides a useful contrast to our recent papers assuming high-precision gravitational wave measurements \cite{Li:2024imk, hybridPrecision}. To determine the effects of using this less certain data, we compare the results of using all the NICER data in Tab. \ref{tab:nicerdata} with two additional calculations. One, we use only PSR J0740+6620 and PSR J0030+0451, whose results have been more consistent across analyses over many years \cite{Riley:2021pdl, Miller:2021qha, Dittmann:2024mbo, Riley:2019yda, Miller:2019cac}. Two, we just remove PSR J0614+3329, which is an outlier for the value of $R_{1.4}$.

To implement the bivariate likelihood, we approximate the asymmetric NICER confidence intervals as Gaussian by averaging the upper and lower 68\% bounds to define $\sigma_M$ and $\sigma_R$. 
The correlation coefficients $\rho_{MR}$ are estimated using the Pearson correlation of the equal-weight posterior samples reported in each analysis.

Posterior probability distributions are sampled using the Metropolis--Hastings algorithm within a Markov Chain Monte Carlo framework. Each chain is evolved for 30,000 burn-in steps, which are discarded, followed by 300,000 production steps used for inference (100,000 steps when only the filter is applied, owing to faster convergence). Most analyses employ eight independent walkers; for the hypothetical high-precision $R_{1.4}$ scenario, we use sixteen walkers to compensate for the lower acceptance rate. Convergence is verified by ensuring that the average accepted parameter values stabilize before the end of burn-in, indicating sampling near equilibrium. Additional methodological details can be found in Ref.~\cite{Li:2024imk}.

\section{Results and Discussions}\label{results}

\subsection{Constraints on the EOS Parameters}
\begin{figure}
    \centering
    \includegraphics[width=\linewidth]{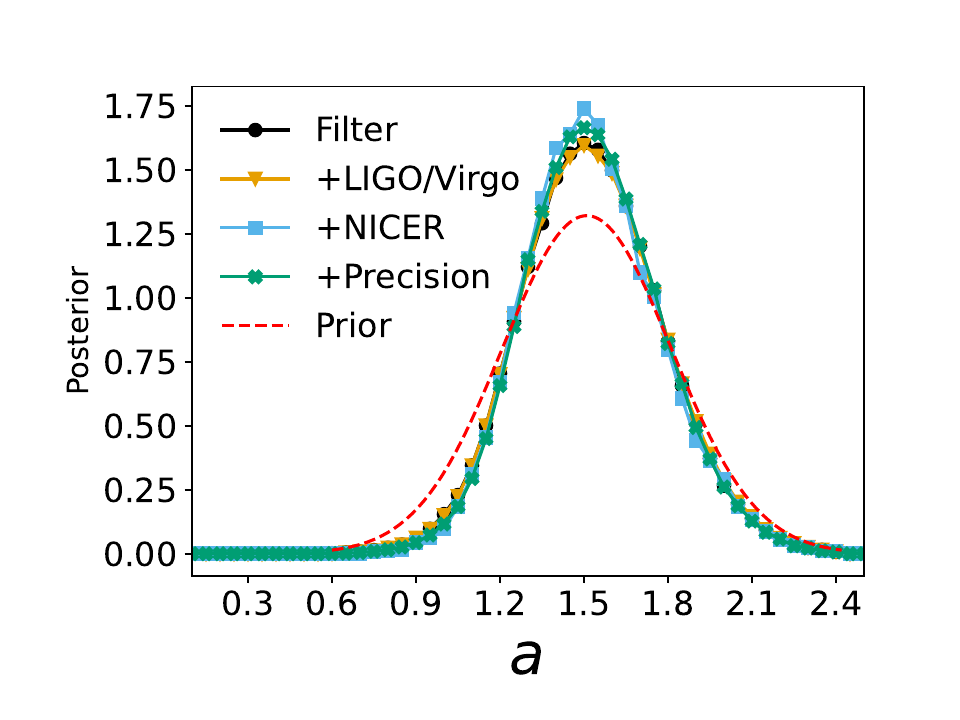}

    \vspace{-1cm}

    \includegraphics[width=\linewidth]{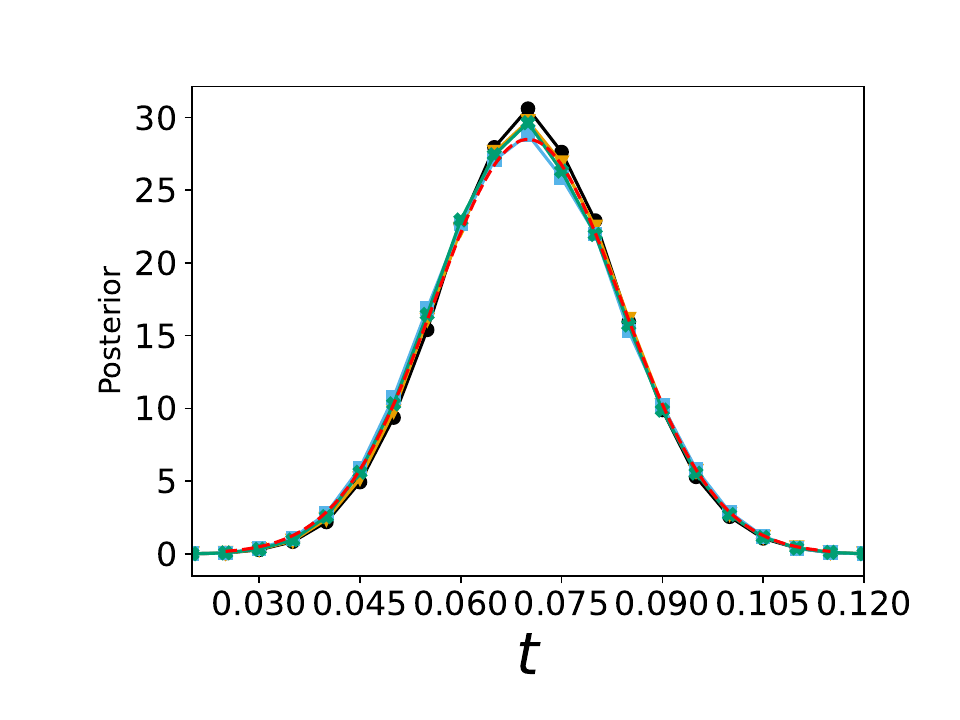}

    \vspace{-0.5cm}
    \caption{PDFs of the QM parameters.}
    \label{fig:qmpara}
\end{figure}

\begin{figure}
    \centering
    \includegraphics[width=\linewidth]{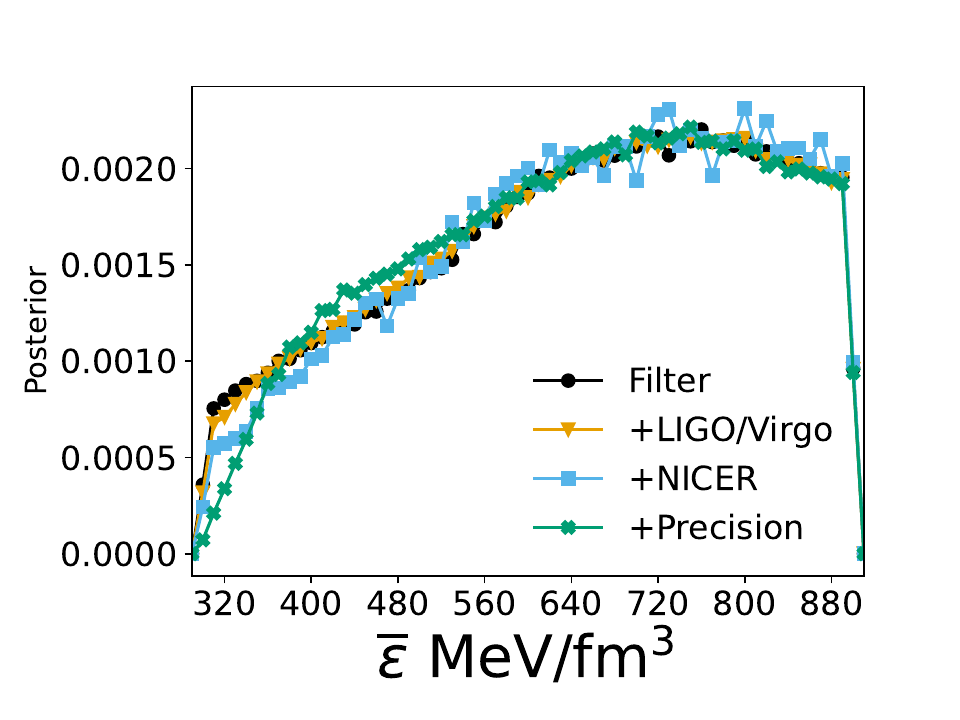}

    \vspace{-1cm}

    \includegraphics[width=\linewidth]{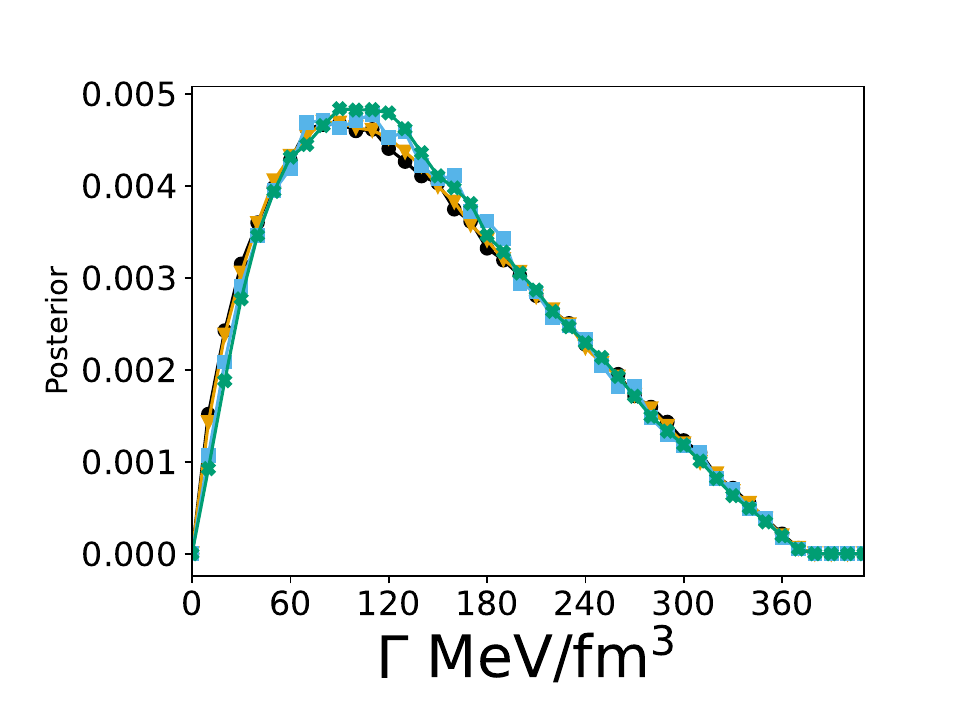}

    \vspace{-0.5cm}
    \caption{Posterior PDFs of the hadron-quark crossover parameters.}
    \label{fig:crosspara}
\end{figure}

Shown in Figs. \ref{fig:hmpara}--\ref{fig:crosspara} are the marginalized 1-D posterior distribution functions (PDFs) of the EOS parameters.

Beginning with the HM parameters in Fig.~\ref{fig:hmpara}, we note that two of them are effectively unconstrained by the data. The posterior PDFs of $K_0$ and $E_0(\rho_0)$ remain essentially identical to their priors. The highest-density parameters for both symmetric nuclear matter and the symmetry energy, $J_0$ and $J_{\rm sym}$, are only weakly constrained by the available observations. While very small values of $J_0$ are disfavored, the posterior distribution exhibits a very broad peak and remains nearly flat toward large values. Since the PDFs of $J_0$ are almost identical for all likelihood functions considered, we conclude that current radius data provide very limited constraints on $J_0$. This conclusion is consistent with results obtained from directly inverting neutron star observables in the $J_0$--$J_{\rm sym}$--$K_{\rm sym}$ parameter space \cite{Zhang:2021xdt}.

Physically, $J_0$ is bounded from above by causality and from below by the requirement that the EOS supports a minimum maximum mass of $M_{\rm TOV} \gtrsim 1.97~M_\odot$ \cite{Zhang:2018bwq}. Since causality provides only a very loose upper bound and realistic EOSs lie well below it, neither the physical filter nor the current radius data significantly constrain the large-$J_0$ region. On the other hand, the suppression of small $J_0$ values arises from their direct impact on $M_{\rm TOV}$, as overly soft EOSs are excluded by the maximum-mass filter.

In the bottom two panels of Fig.~\ref{fig:hmpara}, we observe substantial differences in the posterior PDFs of $L$ and $K_{\rm sym}$ depending on the astrophysical data included. A modest shift toward softer symmetry energy occurs when the LIGO/Virgo radius information is added to the basic physical filter. This is because, in the absence of explicit radius constraints, the requirement $M_{\rm TOV} > 1.97~M_\odot$ favors relatively stiff EOSs. It is well known that the radii of canonical neutron stars are primarily determined by the pressure around $2\rho_0$ \cite{lattimer:2006xb}, where the stiffness of matter depends not only on $J_0$ but also sensitively on $L$ and $K_{\rm sym}$. The inclusion of LIGO/Virgo data therefore favors slightly softer symmetry energy.

\begin{figure}
    \centering
    \includegraphics[width=\linewidth]{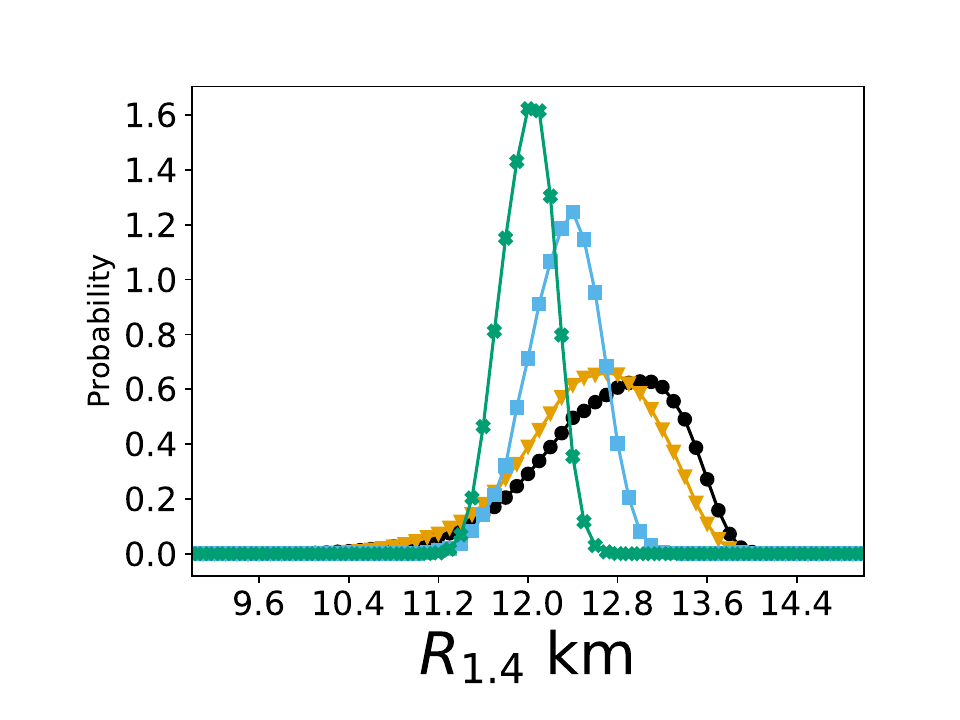}

    \vspace{-1cm}

    \includegraphics[width=\linewidth]{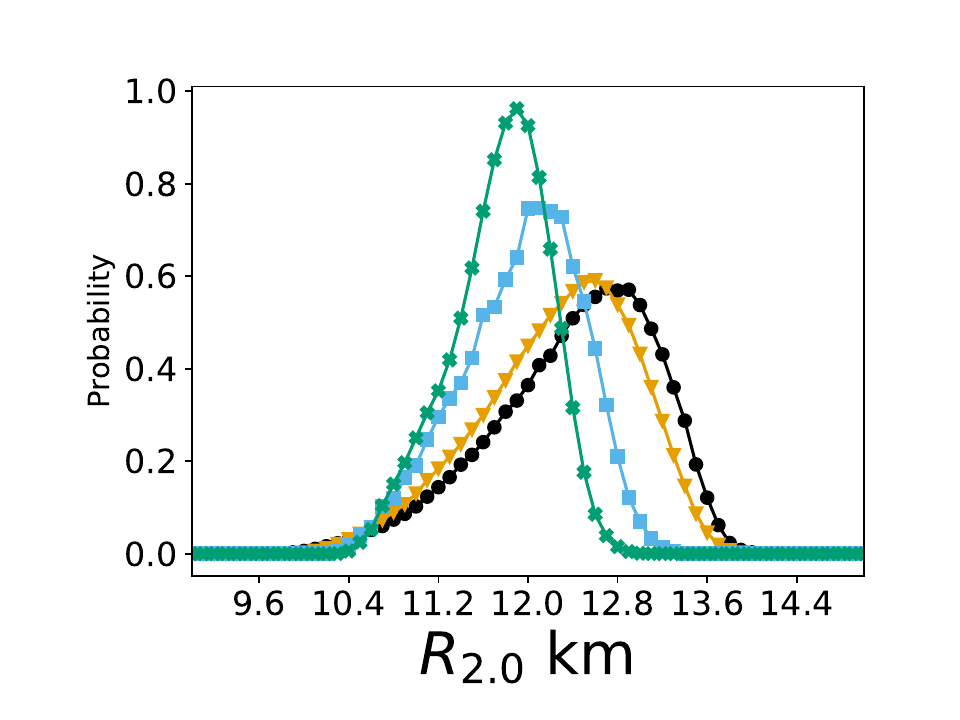}

    \vspace{-1cm}

    \includegraphics[width=\linewidth]{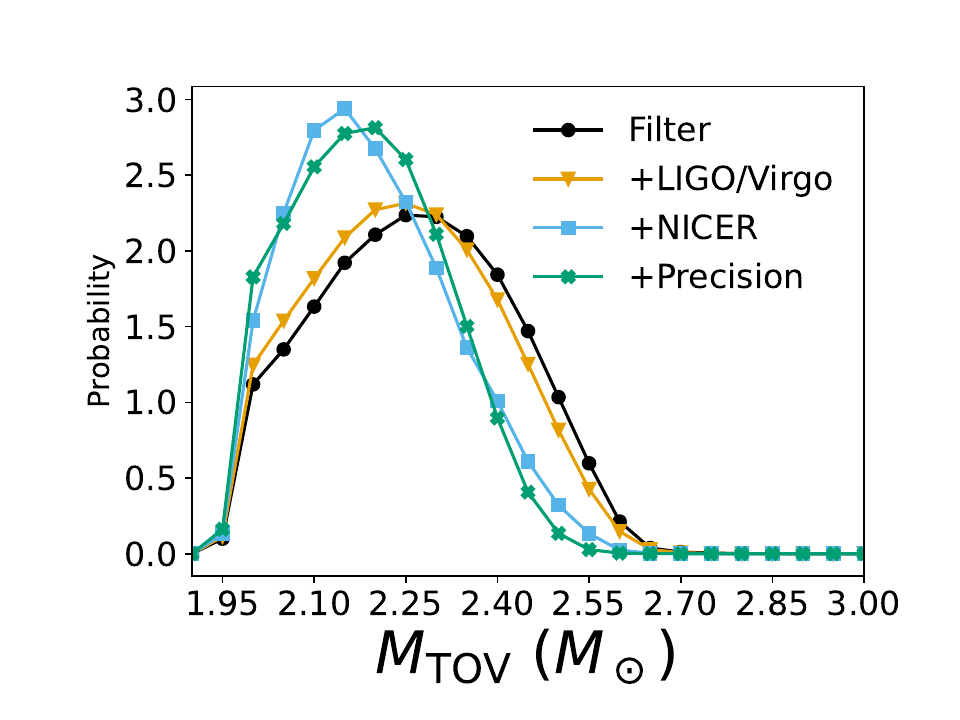}

    \vspace{-0.5cm}
    
    \caption{Posterior probability distributions for NS observables $R_{1.4}$ (upper), $R_{2.0}$ (middle), and $M_{\rm TOV}$ (bottom).}
    \label{fig:nsobs}
\end{figure}

The shift toward smaller values of $L$ and $K_{\rm sym}$ becomes more pronounced when the most recent NICER measurements are used. While PSR J0030+0451 and PSR J0437+4715 are consistent with the LIGO/Virgo results, PSR J0614+3329 exhibits a significantly smaller radius at a comparable mass, implying a softer EOS. Since $R_{1.4}$ is most sensitive to $L$ and $K_{\rm sym}$ \cite{Richter:2023zec}, it is expected that these parameters are most strongly constrained by radius data. The most dramatic shift occurs when we assume a future high-precision measurement of $R_{1.4} = 11.9 \pm 0.2$ km. In this case, the posterior of $L$ shifts toward its lower bound, and the PDF of $K_{\rm sym}$ develops a dominant peak below $-200$~MeV, with a secondary peak at larger values. This behavior was also observed in Ref.~\cite{Li:2024imk} using mock radius data motivated by proposed next-generation X-ray and gravitational-wave observations \cite{Chatziioannou:2021tdi, Pacilio:2021jmq, Bandopadhyay:2024zrr, Finstad:2022oni, Walker:2024loo}. The systematic shift toward softer EOSs arises from the highly nonlinear mapping between the EOS and mass--radius relations through the TOV equations \cite{Li:2024imk}. This feature is further verified in Fig.~\ref{fig:nsobs} by examining the posterior PDFs of $R_{1.4}$, $R_{2.0}$, and $M_{\rm TOV}$. The bimodal structure in the PDF of $K_{\rm sym}$ is a consequence of its correlation with both $L$ and $J_{\rm sym}$ \cite{Li:2024imk}.

In Fig.~\ref{fig:qmpara}, we show the posterior PDFs of the QM (also trace anomaly) parameters together with their assumed priors. The parameter $t$ remains fully consistent with its Gaussian prior, while $a$ exhibits only a marginal deviation. The resulting posteriors are nearly identical across all likelihood functions considered, indicating that current neutron star mass--radius data provide very limited constraints on the QM sector. This implies that present observations do not probe sufficiently deep into the quark-dominated core to extract meaningful information about QM properties, as already pointed out in Ref.~\cite{hybridPrecision}. Nevertheless, as we shall discuss below, this result reflects the universal behavior of the dense-matter trace anomaly \cite{Li:2026tjj}.

Lastly, in Fig.~\ref{fig:crosspara}, we show the posterior PDFs of the hadron--quark crossover parameters. In contrast to the QM parameters, these distributions differ significantly from their uniform priors. The maximum \emph{a posteriori} values are approximately $\overline{\varepsilon} \simeq 750~\mathrm{MeV/fm^3}$ ($\sim 5\varepsilon_0$) and $\Gamma \simeq 100~\mathrm{MeV/fm^3}$ ($\sim 2\varepsilon_0/3$), corresponding to a typical crossover region spanning $\sim 550$--$950~\mathrm{MeV/fm^3}$.

\subsection{Typical Global Observables}

In Fig. \ref{fig:nsobs}, we show the posterior probability distributions of three NS global observables: $R_{1.4}$, $R_{2.0}$, and $M_{\rm TOV}$. If we look at the predicted radii constrained only by basic physics principles included in the filter, we see that $11 \text{ km} \lesssim R_{1.4} \lesssim 14 \text{ km}$ and $10 \text{ km} \lesssim R_{2.0} \lesssim 14 \text{ km}$. The LIGO/Virgo data from GW170817 only slightly favors smaller radii, which is a nontrivial result. While the extracted $R_{1.4}$ value for that event was $11.9 \pm 0.875$ km, our Bayesian inference finds that the most probable value for $R_{1.4}$ is around 12.5--12.8 km, still consistent with measurement uncertainty. If we increase the hypothetical precision of the measurement, however, we force $R_{1.4} = 11.9$, requiring a softer EOS as seen in Fig. \ref{fig:hmpara}. Using the NICER-informed likelihood, we see $11.5 \text{ km} \lesssim R_{1.4} \lesssim 13.0 \text{ km}$ and $10.5 \text{ km} \lesssim R_{2.0} \lesssim 13 \text{ km}$. This scenario represents the most comprehensive current set of neutron star mass–radius constraints.
\begin{figure}
    \centering
    \includegraphics[width=\linewidth]{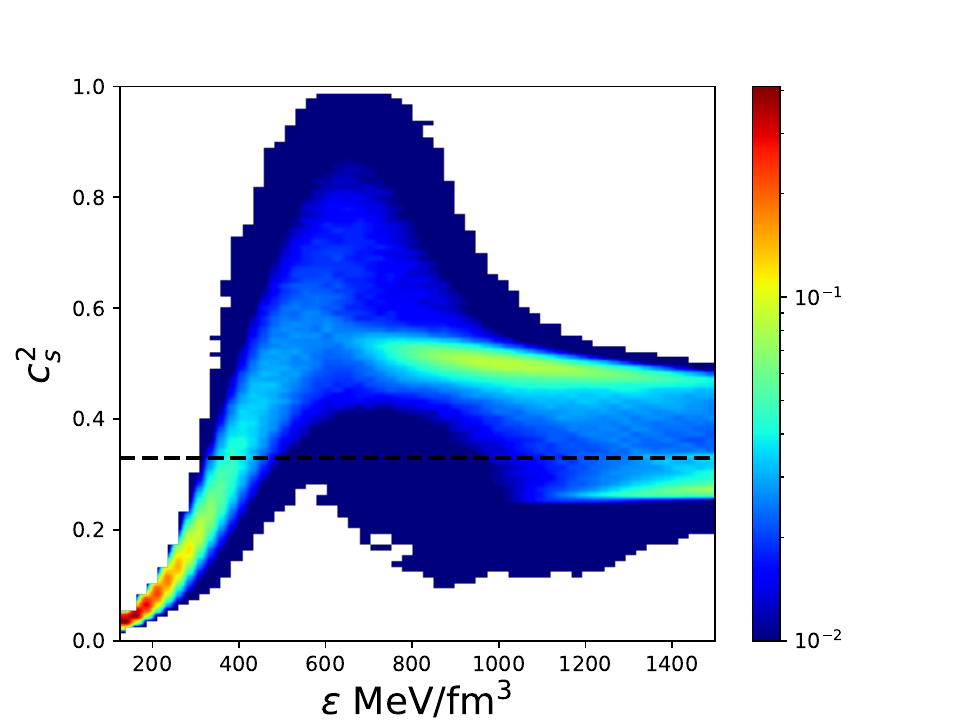}

    \vspace{-0.5cm}

    \includegraphics[width=\linewidth]{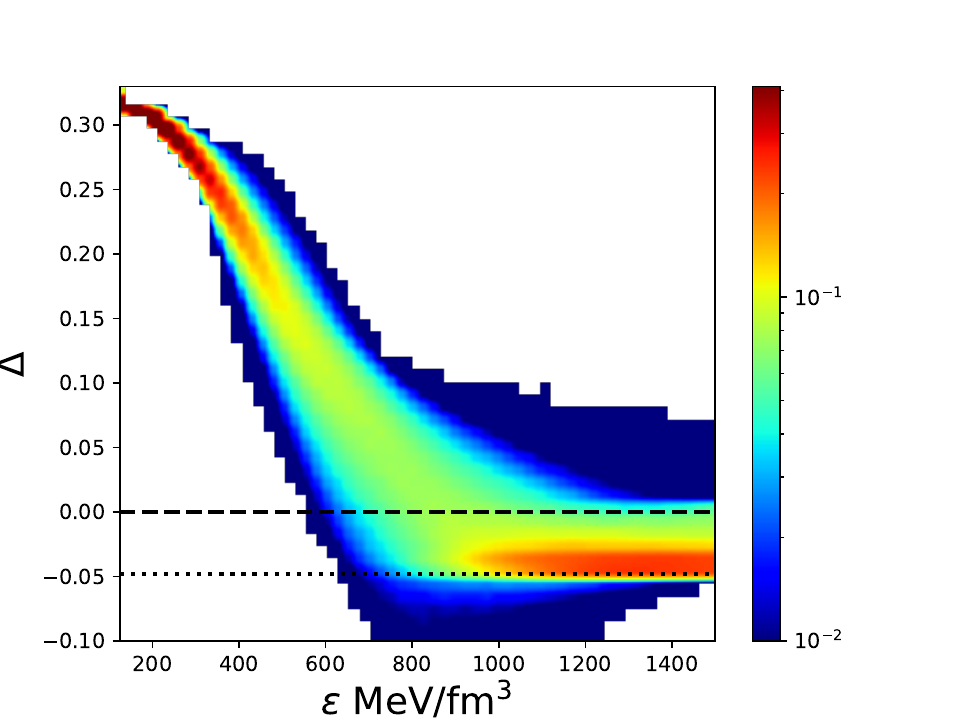}

    \caption{The speed of sound squared and trace anomaly profiles with respect to energy density for accepted EOS using the most recent NICER data. The pQCD conformal limits are indicated by the dashed lines, while the GR limit for the trace anomaly is shown with the dotted line. Every bin was divided by the total number of accepted EOS. Note the logarithmic color scale.}
    \label{fig:c2sTraN_nevent}
\end{figure}
\begin{figure}
    \centering
    \includegraphics[width=\linewidth]{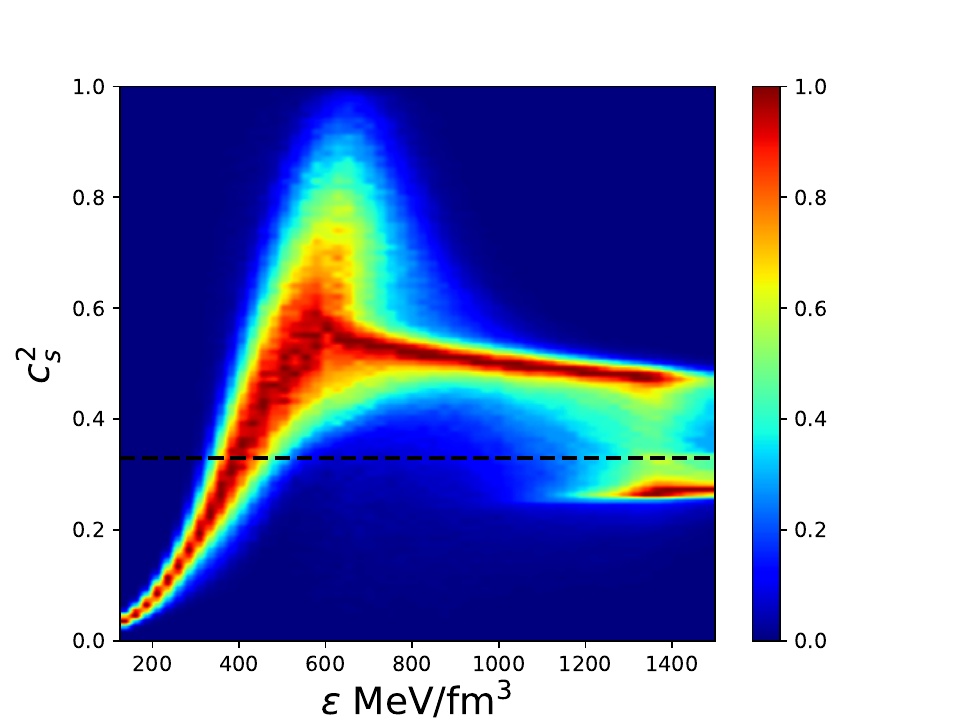}

    \vspace{-0.5cm}

    \includegraphics[width=\linewidth]{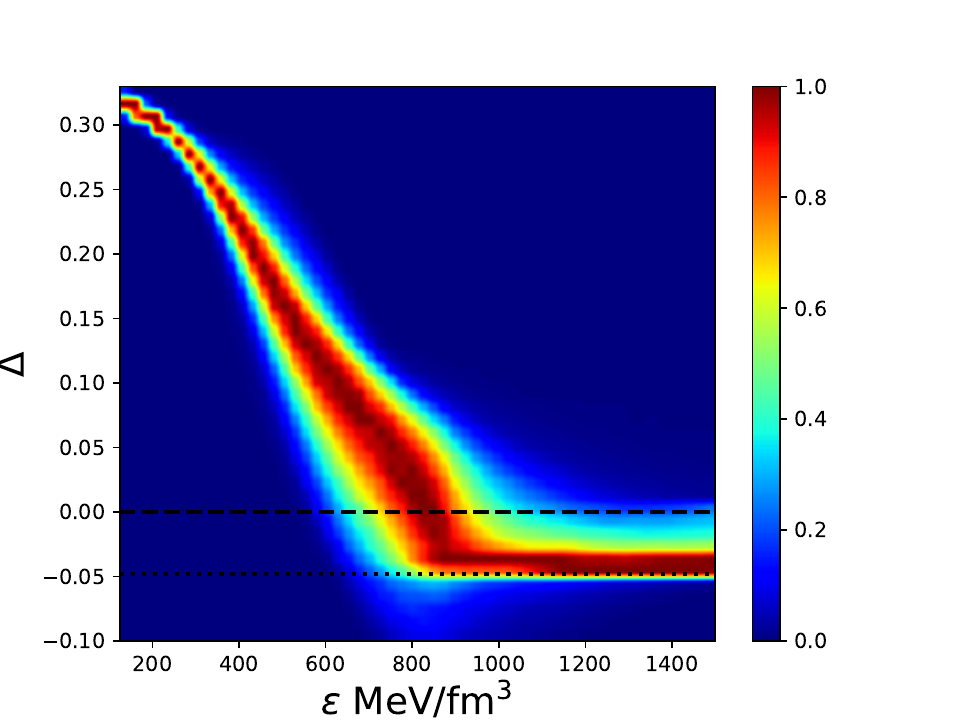}

    \caption{The same as in Fig. \ref{fig:c2sTraN_nevent} but now the count in each energy density bin was scaled to have a maximum value of one.}
    \label{fig:c2sTraN_maxOne}
\end{figure}

\begin{figure}
    \centering
    \includegraphics[width=\linewidth]{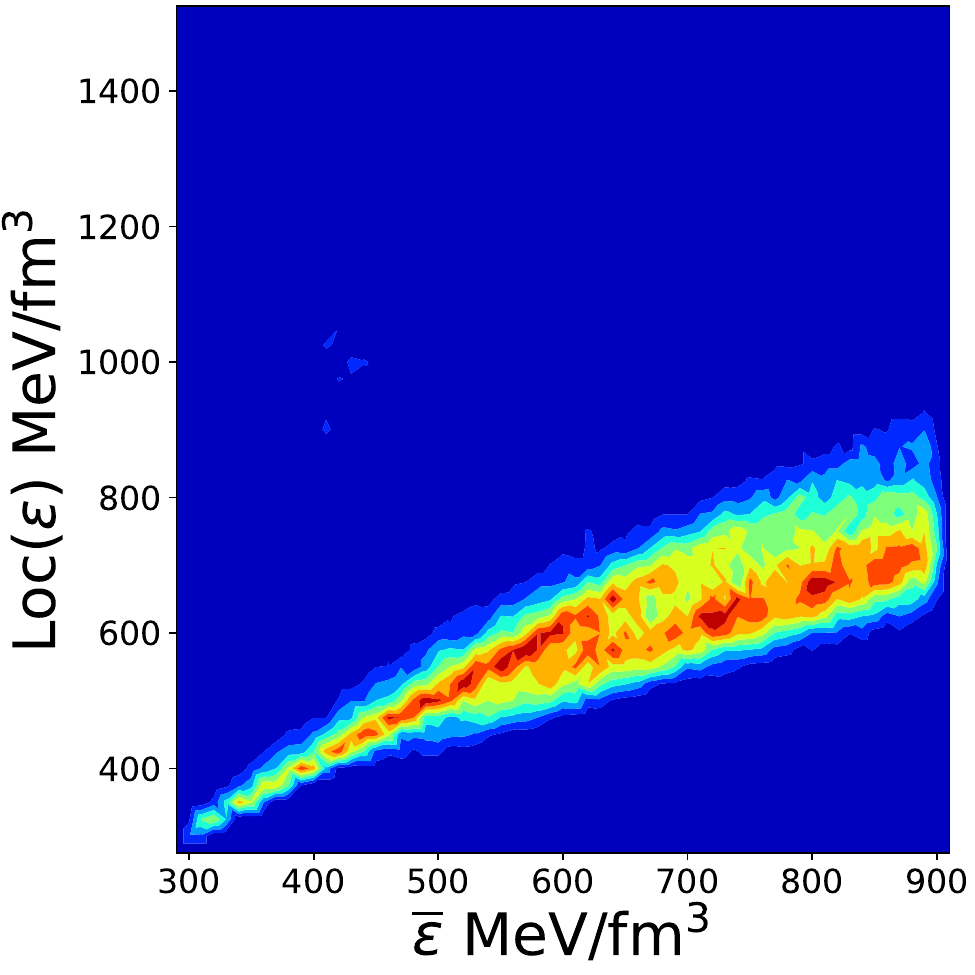}
    \caption{Distribution of pairwise correlation between the central energy density $\overline{\varepsilon}$ of the crossover region and the location Loc$(\varepsilon)$ in energy density of the maximum speed of sound.}
    \label{fig:erhobarloce}
\end{figure}

\begin{figure*}
    \centering
    \addtolength{\tabcolsep}{-1.4em}
    \begin{tabular}{ccc}
        \includegraphics[width=0.37\linewidth]{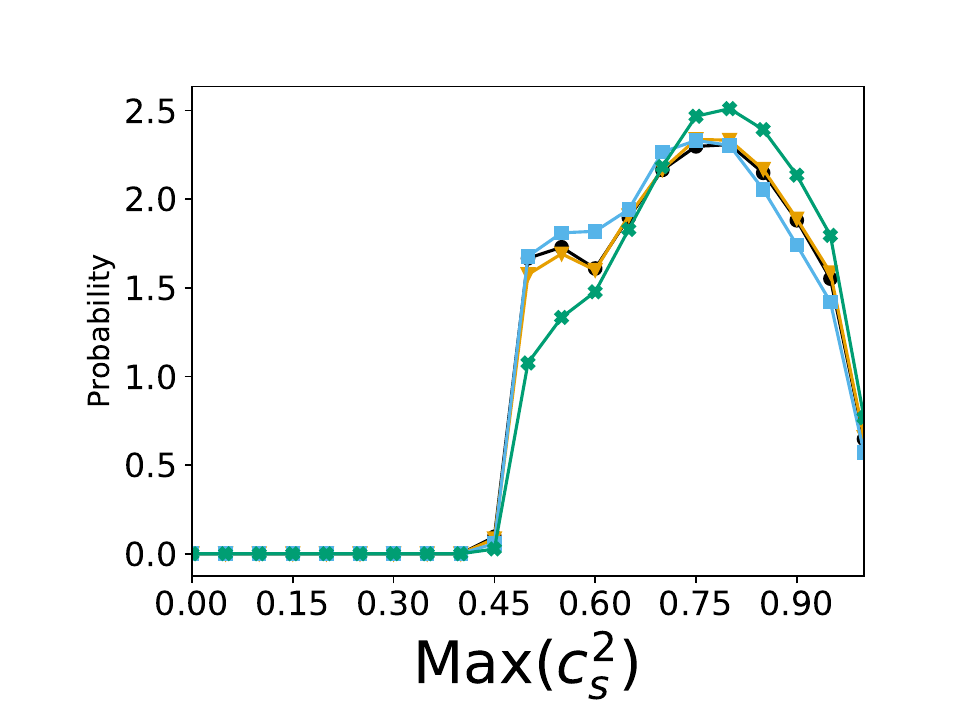} & \includegraphics[width=0.37\linewidth]{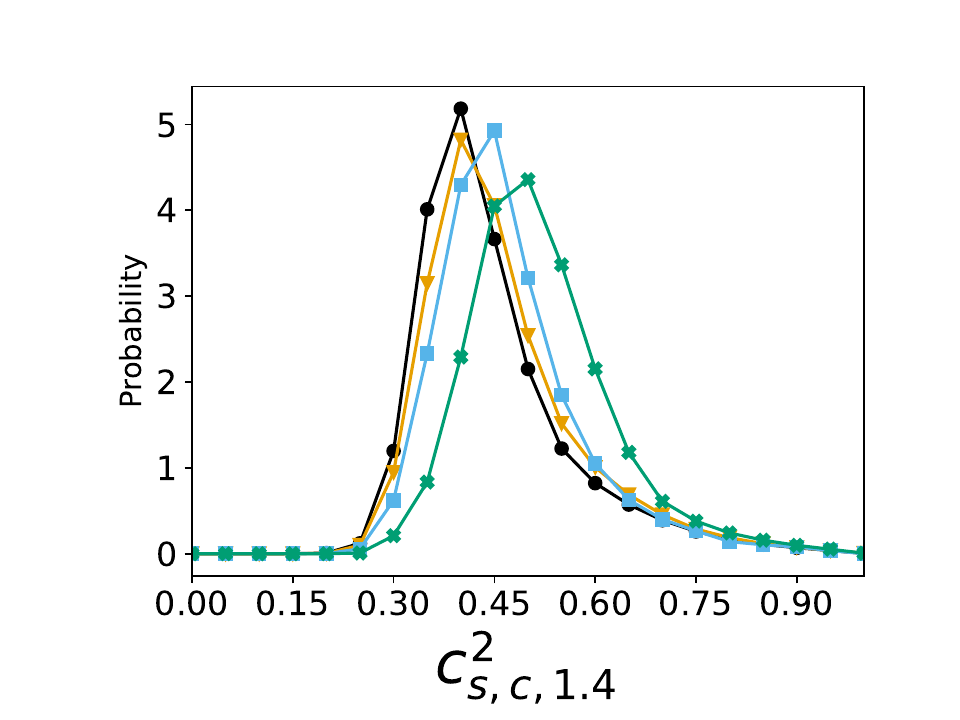} & \includegraphics[width=0.37\linewidth]{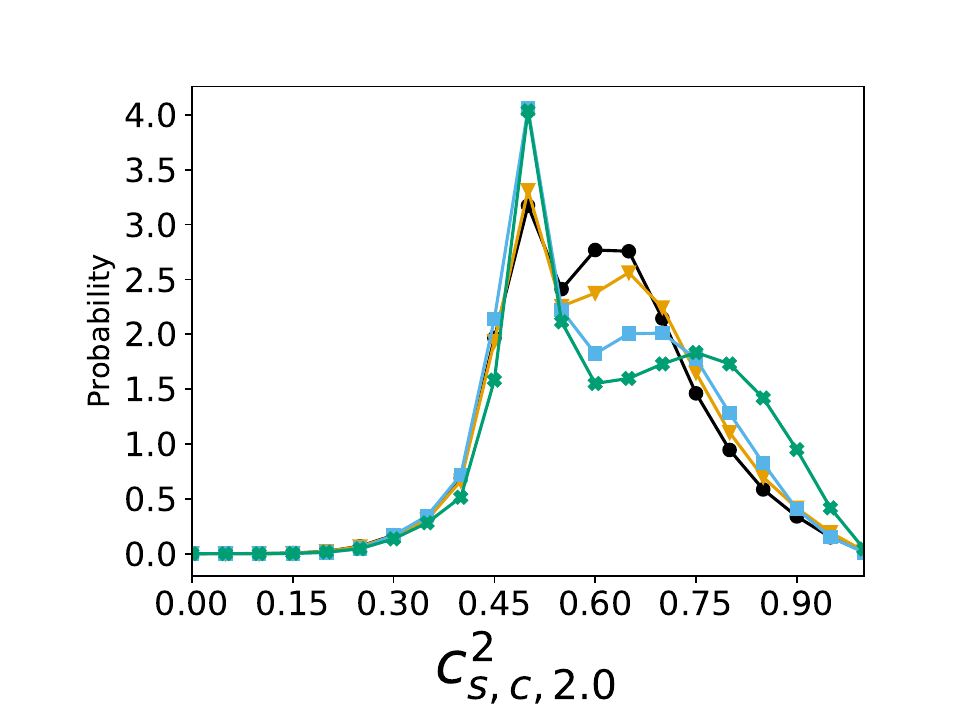}\\[-15pt]
        \includegraphics[width=0.37\linewidth]{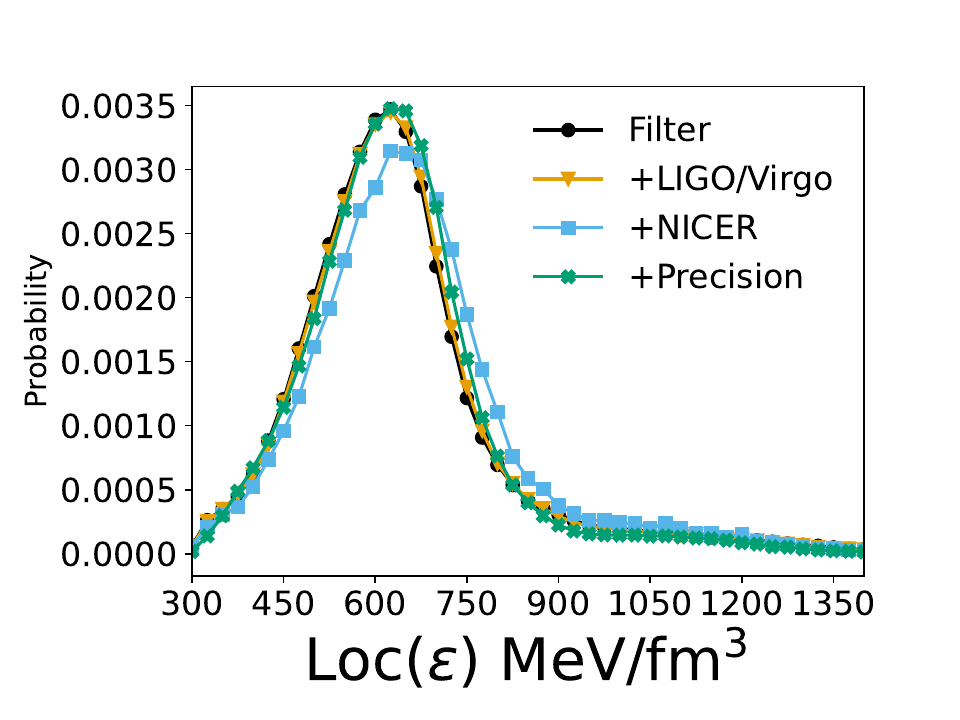} & \includegraphics[width=0.37\linewidth]{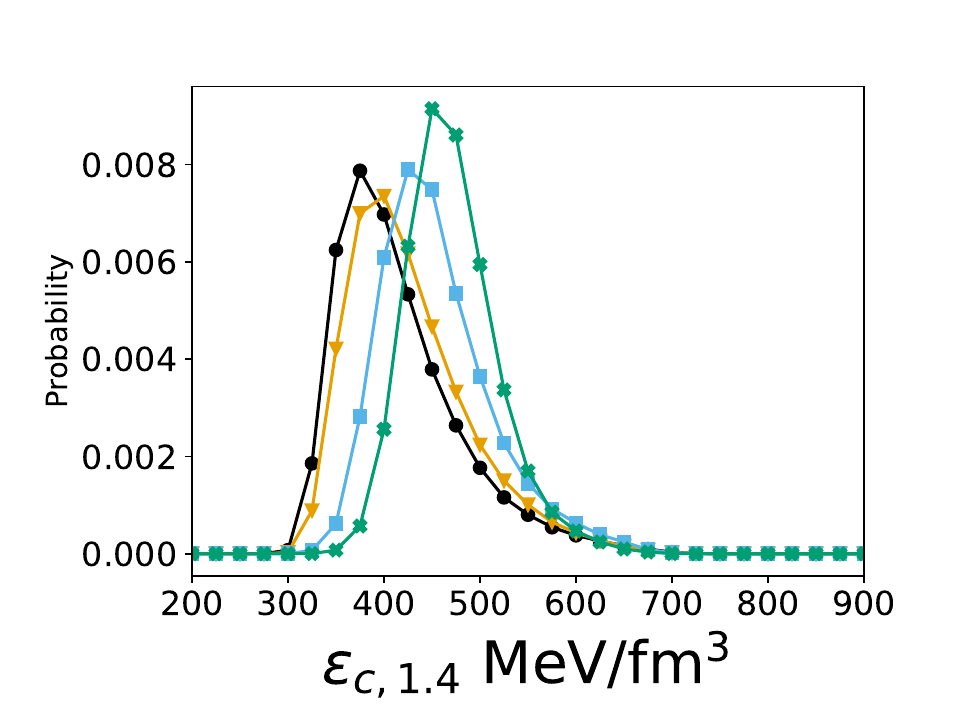} & \includegraphics[width=0.37\linewidth]{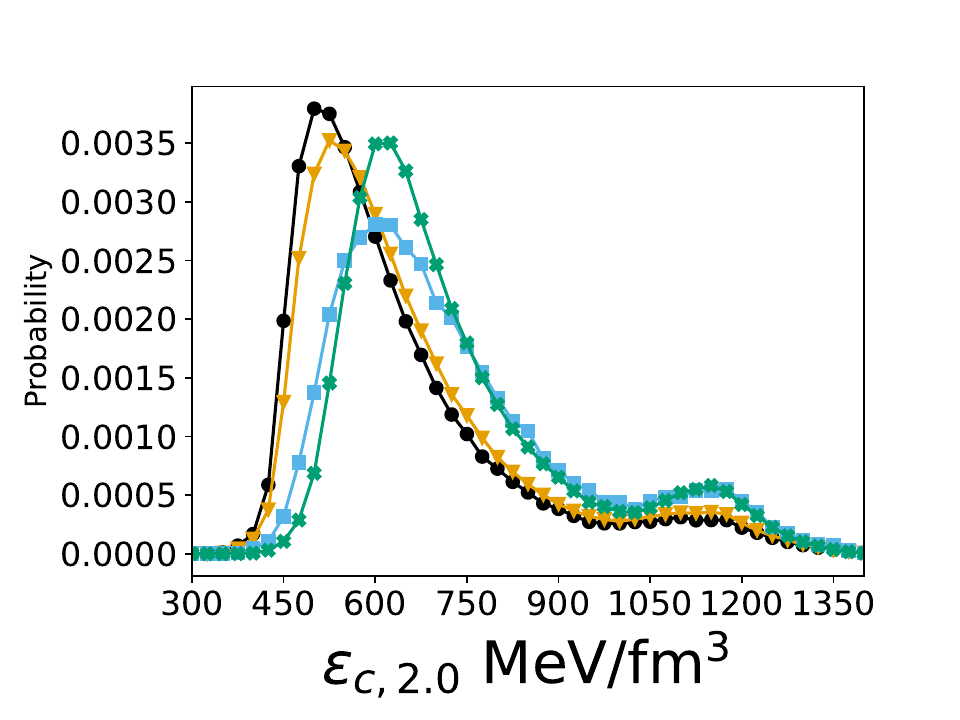}% \\[-15pt]
    \end{tabular}
    \caption{On the left is the probability distribution for the maximum speed of sound squared, Max($c^2_s$), achieved by an EOS (upper panel), and at what energy density, Loc($\varepsilon$), it occurs (lower panel). Also shown are the central energy density and speed of sound in canonical neutron stars (middle) and 2.0 $M_\odot$ NS (right), respectively. The color and symbol codes are the same as in previous figures.}
    \label{fig:nscenter}
\end{figure*}
Looking now at the NS maximum mass $M_{\rm TOV}$ supported by a given EOS, we see that every likelihood used favors $M_{\rm TOV} \approx 2.2~M_\odot$ with slight shifts depending on the exact data used. Notably, the NICER scenario, despite being the only likelihood informed by a nearly 2.1 $M_\odot$ NS, does not predict the most massive NS. This behavior reflects the fact that different neutron star observables probe different density regimes of the EOS. While $R_{1.4}$ is primarily sensitive to the pressure around $2\rho_0$, the maximum mass $M_{\rm TOV}$ is determined by the EOS at much higher densities, typically $\gtrsim 4\rho_0$. As a result, an EOS can be relatively soft at intermediate densities, yielding small radii, yet stiffen rapidly at higher densities and support a large maximum mass. In the filter and LIGO/Virgo scenarios, the relatively broad radius constraints still allow such EOSs, leading to posterior distributions of $M_{\rm TOV}$ extending up to $M_{\rm TOV}\sim2.6\,M_\odot$. In contrast, the NICER likelihood constrains radii at multiple masses with smaller uncertainties, effectively requiring the EOS to remain soft over a wider density range. This suppresses the possibility of rapid high-density stiffening and consequently leads to lower inferred values of $M_{\rm TOV}$, despite the inclusion of a $\sim2.1\,M_\odot$ pulsar in the NICER dataset.

\subsection{Energy-Density Profiles of Speed of Sound and Trace Anomaly}

We now turn to the behavior of the speed of sound and the trace anomaly in neutron stars as functions of the energy density. In the upper panel of Fig.~\ref{fig:c2sTraN_nevent}, we show the squared speed-of-sound profile 
$P(\varepsilon,c_s^2)=N(\varepsilon,c_s^2)/N_{\rm EOS}$, where $N(\varepsilon,c_s^2)$ is the number of occurrences and $N_{\rm EOS}$ is the total number of EOSs accepted in our Bayesian analysis using the most recent NICER dataset. This quantity represents the probability density of finding matter with a given $c_s^2$ at a random point inside neutron stars drawn from the accepted EOS ensemble. The lower panel shows the corresponding joint distribution of the trace anomaly $P(\varepsilon,\Delta)$, defined analogously.

These heat maps are joint measures that combine several effects: how frequently a given $c_s^2(\varepsilon)$ or $\Delta(\varepsilon)$ occurs among the EOS ensemble, how much stellar volume is associated with a given energy density, and how many EOSs allow that energy density at all. First of all, it is not surprising that there is a high probability in the low energy density region, but generally lower probabilities at higher energy densities. This is simply because all EOSs selected by our Bayesian analysis have to go through the low energy density region, but not all of them can get into the high energy density regions, only found in the cores of massive NSs. 
For example, the high probability density at low energy densities, $\varepsilon \lesssim 300~\mathrm{MeV/fm}^3$, which mainly corresponds to the outer regions of all neutron stars, arises from the geometrical $4\pi r^2$ weighting entering the mass integral used during the Bayesian analyses. Toward higher densities in the stellar interior, $c_s^2(\varepsilon)$ generally increases, but the available volume becomes smaller, and few EOSs can support massive NSs where these high energy densities are reached, leading to a reduced probability density.

Interestingly, for EOSs whose central densities lie deep in the post-maximum region of the sound-speed profile, where the EOS undergoes a partial softening associated with the hadron--quark crossover, an extended high-probability region emerges around $c_s^2 \simeq 0.5$. In addition, a smaller secondary peak appears in the joint distribution near $c_s^2 \simeq 0.25$ and $\varepsilon \simeq 1400~\mathrm{MeV/fm^3}$. For these EOSs, the speed of sound decreases after reaching its maximum and either settles into a lower plateau or asymptotically approaches the general-relativistic bound on the trace anomaly, as shown in the lower panel of Fig.~\ref{fig:c2sTraN_nevent}. 

These secondary structures indicate that a large fraction of the accepted EOSs exhibit similar sound-speed behavior around $\varepsilon \sim 800$--$900~\mathrm{MeV/fm^3}$, where they soften toward the quark-matter EOS after the crossover region. This behavior arises because the selected quark-matter EOS is relatively insensitive to the specific parameter choices, leading to a near-universal sound-speed profile in the quark-dominated regime. Consequently, when the stellar central density coincides with these softened segments, a secondary accumulation of probability appears at lower $c_s^2$, a phenomenon that is expected to occur preferentially in massive neutron stars, as we shall demonstrate below.

In addition to the globally normalized joint distribution $P(\varepsilon,c_s^2)$, we also show in Fig.~\ref{fig:c2sTraN_maxOne} an energy-slice--rescaled map in which the maximum count in each $\varepsilon$ bin is normalized to unity. This representation does not correspond to a probability density, but rather serves as a contrast-enhanced visualization that highlights the dominant sound-speed branches at fixed energy density and emphasizes EOS-induced structures. Comparing Figs.~\ref{fig:c2sTraN_nevent} and \ref{fig:c2sTraN_maxOne}, we see that the geometrical volume effects are largely removed in the latter, revealing more clearly the crossover-induced ridges and the apparent universality of the sound-speed profile.

It is worth noting that the trace anomaly $\Delta(\varepsilon)$ is confined to a relatively narrow band, whereas the squared speed of sound exhibits much richer structure. This is because
$\Delta(\varepsilon)=\frac{1}{3}-\frac{P}{\varepsilon}=\frac{1}{3}-\langle c_s^2(\varepsilon) \rangle$
measures the energy-density--averaged squared speed of sound $\langle c_s^2(\varepsilon) \rangle$ \cite{Saes24, Marc24},
\begin{equation}
\langle c_s^2(\varepsilon) \rangle = \frac{1}{\varepsilon}\int_0^{\varepsilon} c_s^2(\varepsilon')\,\dd\varepsilon' 
= \frac{P(\varepsilon)}{\varepsilon}.
\label{eq:PhiAverage}
\end{equation}
As shown earlier in Fig.~\ref{fig:qmpara}, the posterior PDFs of the trace-anomaly parameters $t$ and $a$ are essentially independent of the datasets used, including the high-precision mock radius data, indicating a universal behavior of the dense-matter trace anomaly. This suggests that the trace anomaly provides a robust, composition-insensitive 
macroscopic descriptor of dense matter, largely independent of the microscopic 
details of the EOS within current observational precision~\cite{Li:2026tjj}. This behavior persists even under hypothetical high-precision radius constraints, indicating that current and near-future observations are largely insensitive to the detailed structure of quark matter.

Finally, to assess whether the nonmonotonic behavior of the speed of sound is genuinely induced by the crossover construction rather than being a fine-tuned model artifact, we show in Fig.~\ref{fig:erhobarloce} the pairwise correlation between the crossover parameter $\overline{\varepsilon}$ and the location $\mathrm{Loc}(\varepsilon)$ in energy density of the maximum speed of sound for the accepted EOSs. We find that $\mathrm{Loc}(\varepsilon)$ exhibits a strong, positive, and nearly linear correlation with $\overline{\varepsilon}$. For $\overline{\varepsilon} \lesssim 600~\mathrm{MeV/fm}^3$, the relation is approximately $\mathrm{Loc}(\varepsilon) \approx \overline{\varepsilon}$, indicating that a smooth crossover naturally induces a peak in the speed of sound within the transition region, consistent with previous studies \cite{Masuda:2012ed, Huang:2022mqp}. For $600 \lesssim \overline{\varepsilon} \lesssim 900~\mathrm{MeV/fm}^3$, the correlation weakens, which can be attributed to the fact that the hadronic EOS itself may generate a peak in $c_s^2(\varepsilon)$, depending sensitively on the high-density behavior of the nuclear symmetry energy, as demonstrated in Refs.~\cite{Zhang:2022sep, Ye:2024meg} and reviewed in Ref.~\cite{Li:2025xio}. In this case, if the crossover sets in at sufficiently high densities, the observed peak in $c_s^2$ may originate primarily from the hadronic sector.

\begin{figure*}
    \includegraphics[width=0.36\linewidth]{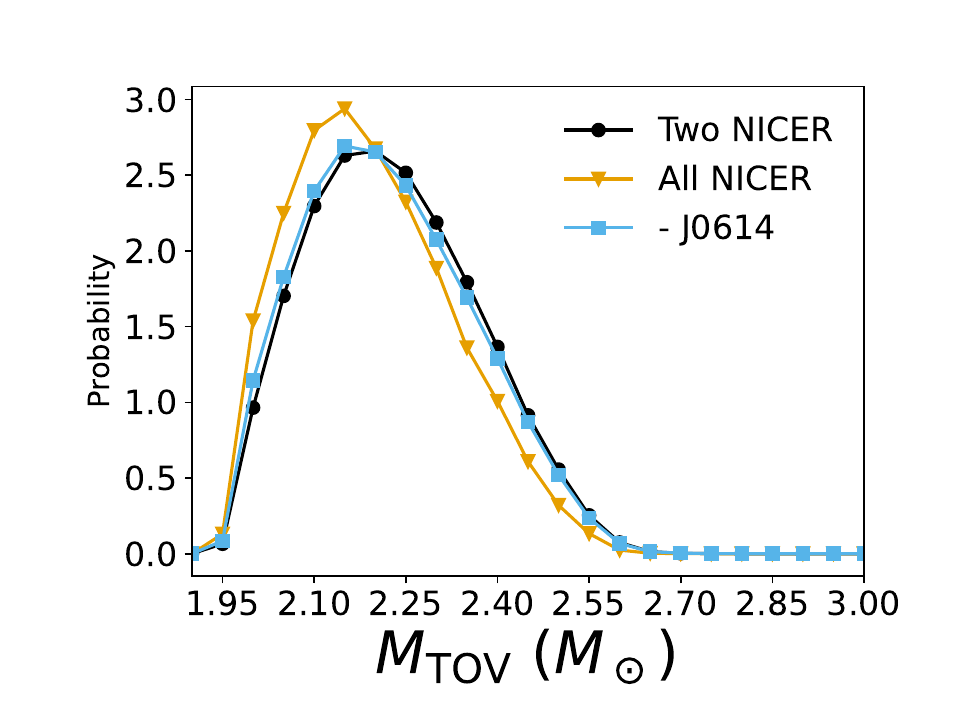} \hspace{-1.1cm} \includegraphics[width=0.36\linewidth]{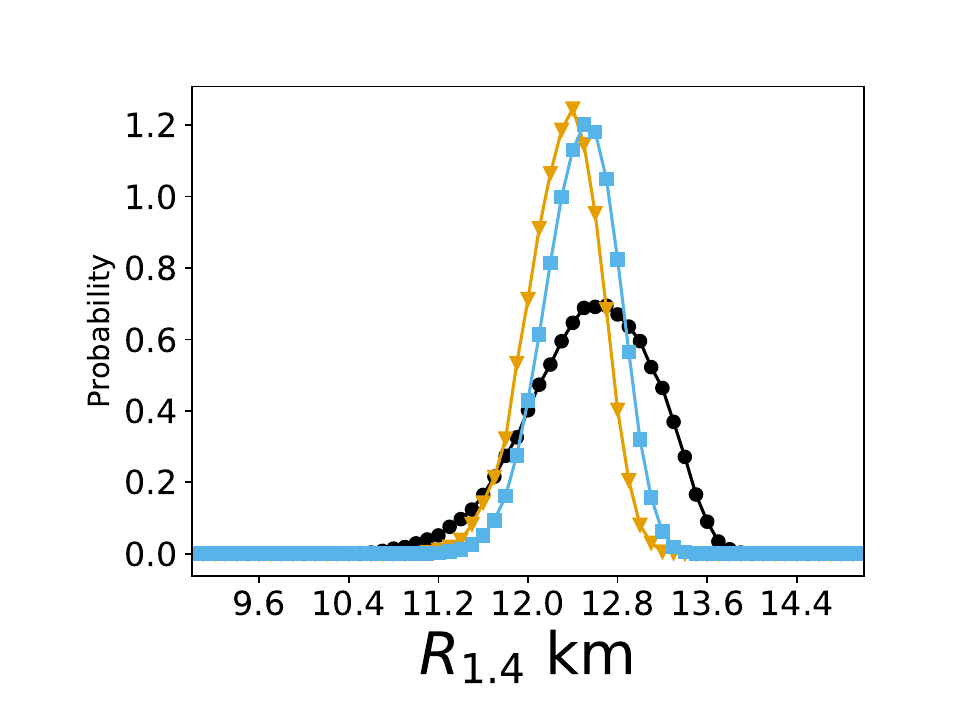} \hspace{-1.1cm} \includegraphics[width=0.36\linewidth]{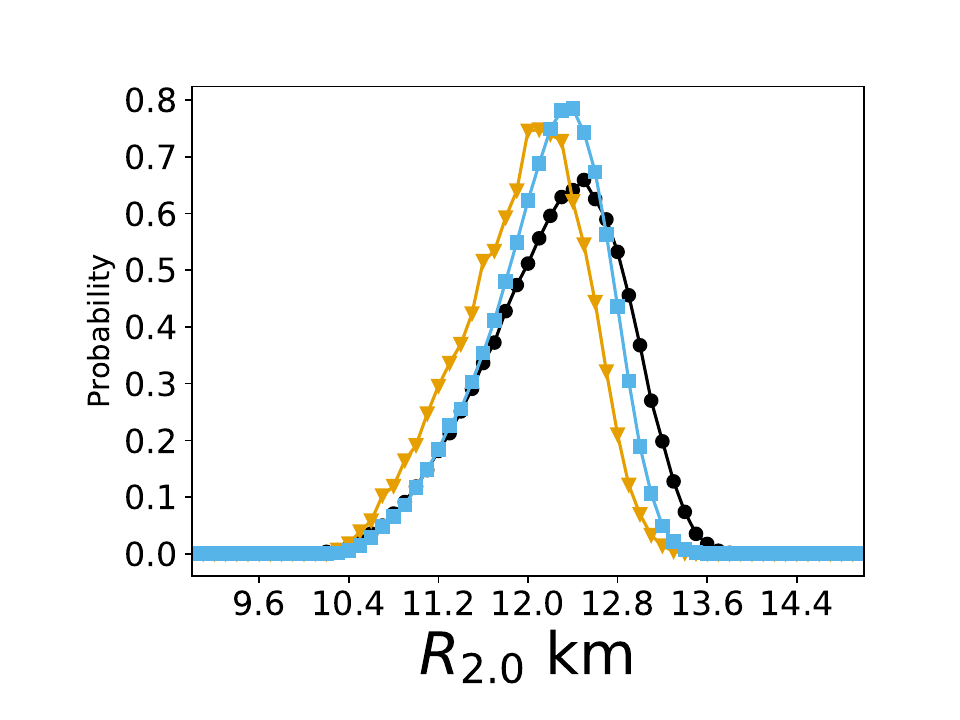}\\
    \includegraphics[width=0.37\linewidth]{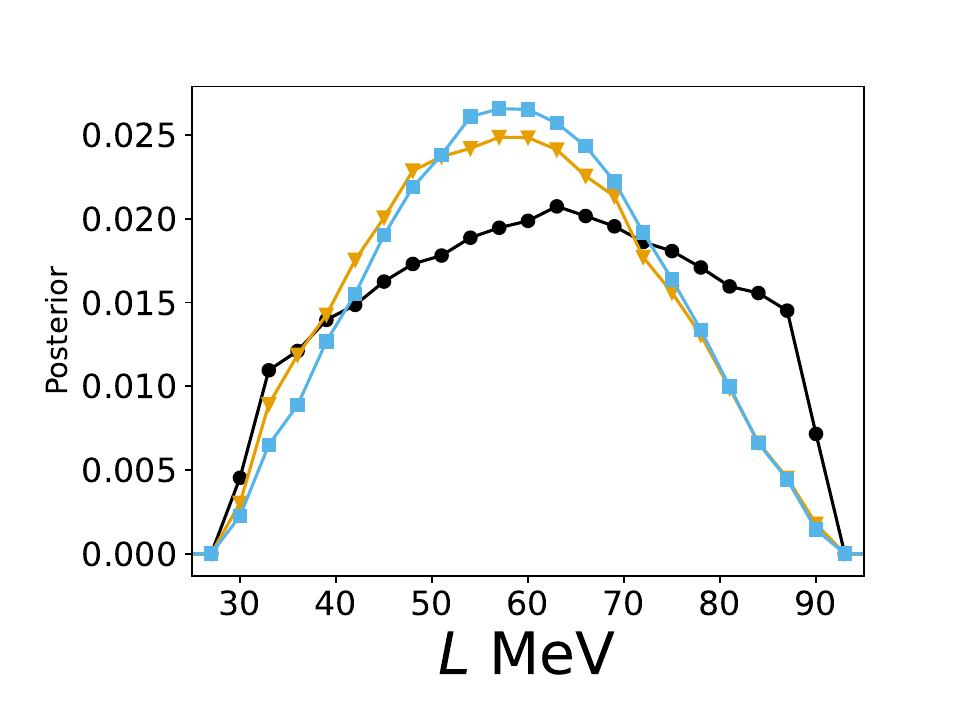} \includegraphics[width=0.37\linewidth]{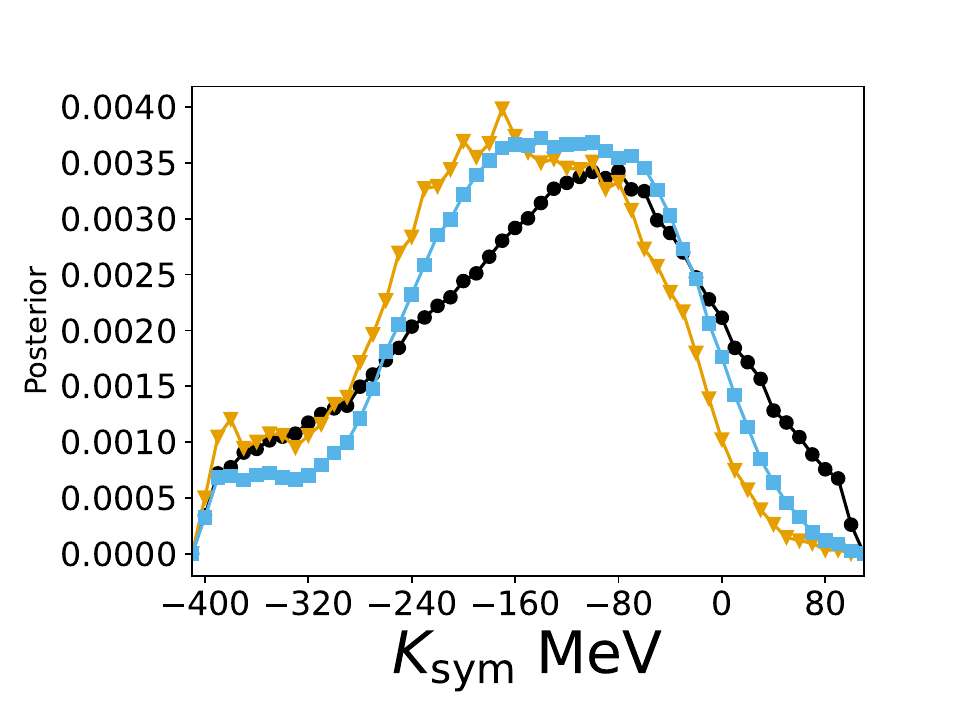}\\
    \includegraphics[width=0.37\linewidth]{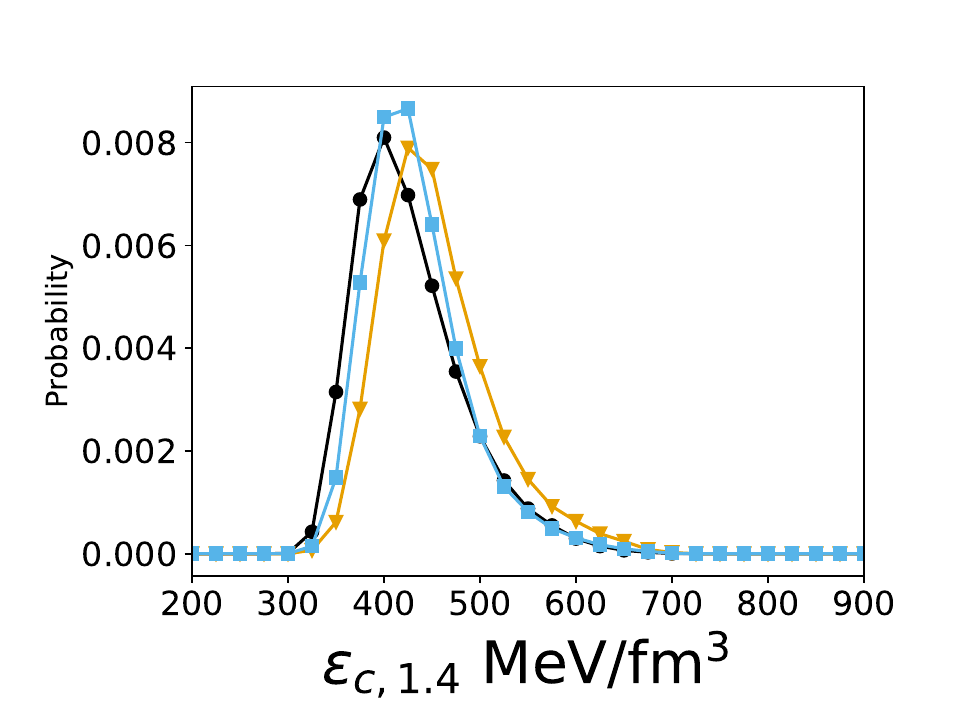} \includegraphics[width=0.37\textwidth]{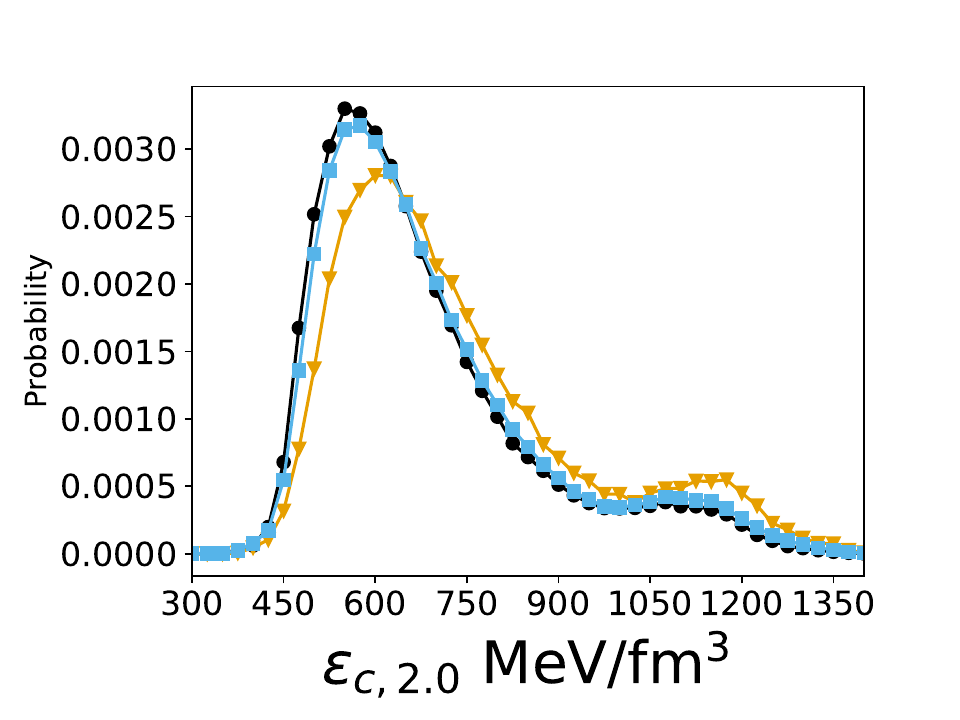}
    \caption{The PDFs of observables, parameters, and NS interiors that differed significantly depending on the NICER data used.}
    \label{fig:nicercomparison}
\end{figure*}

\subsection{Nonmonotonic Speed-of-Sound Profiles in Massive Neutron Stars}
The implications of the features studied above become evident when examining neutron stars of different masses. The top-left panel of Fig.~\ref{fig:nscenter} shows the probability distribution of the maximum squared speed of sound, $\max(c_s^2)$, reached by the accepted EOSs. We find that all accepted EOSs violate the conformal bound $c_s^2 < 1/3$. The bottom-left panel shows the corresponding energy density $\mathrm{Loc}(\varepsilon)$ at which $\max(c_s^2)$ occurs. The most probable value is $\mathrm{Loc}(\varepsilon) \simeq 600~\mathrm{MeV/fm}^3$ ($\sim 4\varepsilon_0$), placing it in the early part of the typical crossover region.

In the middle and right panels of Fig.~\ref{fig:nscenter}, we present the posterior distributions of the central squared speed of sound and central energy density for $1.4~M_\odot$ and $2.0~M_\odot$ neutron stars, respectively. Focusing first on the central energy density, we observe that the most probable value shifts systematically toward higher densities when going from the filter-only case to the LIGO/Virgo, NICER, and precision scenarios. This ordering reflects the corresponding trend from stiffer to softer hadronic EOS parameters, as seen in Fig.~\ref{fig:hmpara}, and is consistent with the well-known result that stiffer EOSs yield lower central densities at fixed stellar mass \cite{lattimer:2006xb, Lattimer:2012nd, Ozel:2016oaf}. Physically, this behavior follows from the hydrostatic balance governed by the TOV equations: a stiffer EOS provides larger pressure at a given energy density, allowing the star to support its mass at lower central densities, whereas a softer EOS requires higher central densities to achieve equilibrium.

For the central speed of sound in a $1.4~M_\odot$ neutron star, the posterior distribution exhibits a single dominant peak at $c_{s,c,1.4}^2 \simeq 0.5$, essentially independent of the likelihood adopted. For a $2.0~M_\odot$ neutron star, all scenarios display a main peak at $c_{s,c,2.0}^2 \simeq 0.5$ together with a pronounced shoulder extending to larger values, whose detailed shape depends mildly on the specific dataset. This high-$c_s^2$ shoulder corresponds to EOSs for which the central density of a $2.0~M_\odot$ star lies close to the density where the speed of sound reaches its maximum.

At the same time, the distribution of $\varepsilon_{c,2.0}$ exhibits a long high-density tail, in which the central density exceeds the location of the peak in $c_s^2$. Beyond this point, the speed of sound decreases with increasing energy density. Comparing $\varepsilon_{c,2.0}$ with the typical crossover region inferred from Fig.~\ref{fig:crosspara}, $\varepsilon \simeq 550$--$950~\mathrm{MeV/fm^3}$, we find that even $2.0~M_\odot$ neutron stars are, in most cases, located only at the onset of the crossover. Consequently, $1.4~M_\odot$ stars experience essentially no quark-matter effects, while $2.0~M_\odot$ stars probe at most the early crossover regime. This explains why the neutron star observations considered here provide little direct sensitivity to the quark-matter EOS and why the quark-sector parameters remain largely consistent with their priors, as seen in Fig.~\ref{fig:qmpara}.

\subsection{Effects of Using Different NS Datasets}

Lastly, we examine the impact of using different subsets of NICER data in constructing the Bayesian likelihood. To avoid redundancy, Fig.~\ref{fig:nicercomparison} displays only those posterior distributions that exhibit significant variations among the scenarios. The first scenario includes only PSR J0740+6620 and PSR J0030+0451; the second uses all NICER data listed in Table~\ref{tab:nicerdata} (which is just the results shown in the previous subsections); and the third includes all NICER data except PSR J0614+3329.

In the first row of Fig.~\ref{fig:nicercomparison}, the inclusion of PSR J0614+3329, which favors a softer EOS due to its smaller inferred radius, does not reduce $M_{\rm TOV}$. This is expected, as the maximum mass is primarily constrained by PSR J0740+6620, the most massive neutron star in our dataset. In contrast, the radius—particularly $R_{1.4}$—becomes more tightly constrained when additional data are included, as reflected by the narrower posterior distributions for the “All NICER” case compared to “Two NICER.” The inclusion of PSR J0614+3329 leads to only a slight reduction in $R_{1.4}$, since other measurements favor larger radii.

The EOS parameters most sensitive to the choice of data are $L$ and $K_{\rm sym}$, consistent with the fact that the additional observations primarily constrain $R_{1.4}$, to which these parameters are most sensitive \cite{Richter:2023zec}. With more data, their posterior distributions shift toward softer values and become moderately more constrained. However, including PSR J0614+3329 does not significantly alter the results relative to the case without it, indicating that these parameters are bounded from below by other observational constraints, such as the maximum mass.

In the third row of Fig.~\ref{fig:nicercomparison}, the central energy densities corresponding to $R_{1.4}$ and $R_{2.0}$ neutron stars are slightly higher when PSR J0614+3329 is included. This reflects the requirement that a softer EOS must reach higher central densities to support the same stellar mass.

\section{Summary and Conclusions}\label{summary}

We have performed a Bayesian inference of the dense-matter equation of state within a unified framework that consistently incorporates hadronic matter, quark matter, and a smooth hadron–quark crossover. By combining physical consistency conditions with gravitational-wave constraints, NICER mass–radius measurements, and hypothetical high-precision observations, we have quantified how current data constrain the properties of dense matter.

Our results show that present observations primarily constrain the low-to-intermediate density regime of the EOS, particularly the slope and curvature of the nuclear symmetry energy. In contrast, the highest-density hadronic parameters, as well as the properties of quark matter and the detailed structure of the crossover, remain only weakly constrained. This reflects the fact that current neutron star observations probe densities up to only a few times nuclear saturation density.

Within this framework, a smooth hadron–quark crossover generically induces a pronounced peak in the speed of sound, whose location correlates strongly with the crossover density. This establishes a direct physical link between the microscopic structure of the EOS and macroscopic neutron star observables. We further find that the trace anomaly exhibits a remarkably universal behavior across the accepted EOS ensemble and remains largely insensitive to current observational constraints. This indicates that the trace anomaly serves as a robust, composition-insensitive macroscopic descriptor of dense matter.

An important outcome of this work is that, even within a flexible crossover framework, current neutron star observations provide only limited sensitivity to the properties of quark matter. Instead, they primarily constrain the macroscopic stiffness of dense matter below $\sim 3$–$4\,\rho_0$. This behavior persists even under hypothetical high-precision radius constraints, indicating that current and near-future observations are largely insensitive to the detailed structure of quark matter.

Finally, by treating hadronic, quark, and crossover sectors on equal statistical footing within a single Bayesian framework, this work provides a unified and model-flexible approach to interpreting neutron star observations across all relevant density regimes.
\\

\textit{Acknowledgement.-}
We thank B.J. Cai, W.J. Xie, and N.B. Zhang for helpful discussions. This work was supported in part by the U.S. Department of Energy, Office of Science, under Award No. DE-SC0013702 and NASA-Texas Space Grant Consortium.

\section*{DATA AVAILABILITY} All data used in this work are publicly available \cite{dataset}.

\bibliographystyle{nst}
\bibliography{refs}

\end{document}